\documentclass[aps,pre,showpacs,twocolumn]{revtex4}
\usepackage{bm}
\usepackage{graphicx}
\bibstyle{approve.bib}
\usepackage{amssymb}
\usepackage{epstopdf}
\usepackage{color}
\newcommand{\be}{\begin{equation}}
\newcommand{\ee}{\end{equation}}
\newcommand{\bea}{\begin{eqnarray}}
\newcommand{\eea}{\end{eqnarray}}
\newcommand{\lan}{\left\langle}
\newcommand{\ran}{\right\rangle}
\newcommand{\br}{\mathbf{r}}
\newcommand{\ba}{\mathbf{a}}

\newcommand{\bom}{\mathbf{\Omega}}
\newcommand{\bk}{\mathbf{k}}

\newcommand{\e}{\varepsilon}
\newcommand{\epa}{\varepsilon_\parallel}
\newcommand{\epe}{\varepsilon_\perp}

\newcommand{\tv}{\tilde{v}}

\newcommand{\pa}{\parallel}

\begin{document}

\title{Dipolar correlations in structured solvents under nanoconfinement}
\author{Sahin Buyukdagli and Ralf Blossey}
\affiliation{Interdisciplinary Research Institute, Universit\'e des Sciences et des Technologies
de Lille (USTL), USR CNRS 3078, 50 Avenue Halley, 59658 Villeneuve d'Ascq, France}

\begin{abstract}
We study electrostatic correlations in structured solvents confined to nanoscale systems. We derive variational equations of Netz-Orland type for
a model liquid composed of finite size dipoles. These equations are solved for both dilute solvents and solvents at physiological concentrations in a slit 
nanopore geometry. Correlation effects are of major importance for the dielectric reduction and anisotropy of the solvent resulting from dipole image 
interactions and also lead to a reduction of van der Waals attractions between low dielectric bodies. Finally, by comparison with other recently 
developed self-consistent theories and experiments, we scrutinize the effect of solvent-membrane interactions on the differential capacitance of the charged 
liquid in contact with low dielectric substrates. The interfacial solvent depletion driven by solvent-image interactions plays the major role in the observed 
low values  of the experimental capacitance data, while non-locality associated with the extended charge structure of solvent molecules only brings a minor 
contribution.
\end{abstract}


\date{\today}
\maketitle

\section{Introduction}

In nanoscale systems governed by electrostatics, it is impossible to overestimate the role of water as a regulator of electrostatic interactions. A well-known example is the stability of charged macromolecules driven by two competing forces. In colloidal systems,  repulsive surface charge interactions are subject to dielectric screening \textcolor{black}{(i.e. the reduction of the surface field by the reaction field induced by solvent molecules)}, and attractive van der Waals (vdW) forces are directly induced by the dielectric contrast between the colloids and the surrounding solvent~\cite{isr}.  Water plays also an essential role in tuning the strength of ion-substrate interactions that drive the selectivity of biological and artificial membrane nanopores~\cite{Hille,Bob,yar1,yar2}, the charge storage ability of capacitor devices~\cite{exp,rev2}, and the efficiency of nanofluidic transport technics~\cite{bocq}. A consistent formulation of the electrostatics of water and mobile ions is thus needed for an analytical insight into the functioning of these systems.

In the standard formulation of macromolecular interactions called the Derjaguin-Landau-Verwey-Overbeek (DLVO) theory~\cite{DLVO,isr}, water is a dielectric continuum liquid whose dielectric response properties are assumed to be unaffected by the presence of the macromolecules. This approximation - made for practical reasons - clearly lacks a solid theoretical basis. Indeed, any improvement over this continuum approach requires an explicit modelling of water electrostatics. The same dielectric continuum approximation is also the basic assumption of ion transport theories that aim at predicting the filtration ability of artificial membrane nanopores used in water purification technics. A big challenge in water desalination technology consists in reducing the high cost of the salt removal process. The current filtration membranes being too impermeable to water, one has to provide important amounts of energy to transfer water through the membrane pores. In order to optimize the permeability of membrane nanopores to water while keeping their ionic filtration efficiency, the task consists in extending our understanding of the solvent-membrane interactions. Because the above-mentioned effects are induced by electrostatic correlations, this requires a solvent-explicit formulation of electrostatic interactions beyond mean-field (MF) level.

Ionic correlation effects in dielectric continuum theories have attracted considerable attention during the last three decades~\cite{PodWKB,attard,netzcoun,netzvar,cyl1,1loop,st3,Lue1,jcp2}. However, theoretical attempts to include solvent  into electrostatics have been mainly limited to MF approaches. These formalisms can be classified into two categories. The first direction corresponds to phenomenological approaches that accounts for the non-local dielectric response of solvent in terms of an effective dielectric permittivity function $\e(\br,\br')$~\cite{Kor1,blossey2}.  The second direction consists of including into the Poisson-Boltzmann (PB) formalism the solvent molecules as simple point-dipoles, a first-order approximation that neglects multipolar and non-local effects. In this category, the first MF level dipolar PB (DPB) formalism was introduced  in Ref.~\cite{dunyuk}, which was later extended by adding steric effects~\cite{orland1} and correlations in bulk solvents~\cite{orland2}. \textcolor{black}{At this point, one should also mention that field-theoretic modified PB theories were, among others, applied to fluctuating polyelectrolytes~\cite{dunpol}, and dipolar~\cite{bohdip} and multipolar molecules~\cite{Kanduc,Lue2}.} In Ref.~\cite{epl}, one of us (SB) included surface polarization effects beyond MF level into the DPB formalism in order to explain the low capacitances of materials with low polarity, leading to the extended-dipolar Poisson-Boltzmann equation (EDPB). Subsequently, in Ref.~\cite{NLPB1} the first microscopic formulation of non-local electrostatics with a structured solvent was introduced, going beyond the point-dipole approximation. We investigated the non-local dielectric response of this polar liquid model in the MF limit of the theory. We reconsidered the same non-local theory in order to shed light on the amplification of bare ionic polarizabilities in bulk solvents~\cite{NLPB2},
and studied its non-local nonlinear response \cite{SBRB}.

The present work aims at elucidating charge fluctuation effects in confined polar liquids. To this end, we reconsider the solvent model of Ref.~\cite{NLPB1} beyond the MF limit. Starting with the field-theoretic representation of the grand-canonical partition function we derive in Section~\ref{genvar} the self-consistent (SC) equations that explicitly include the charge structure of the solvent molecules, here taken as simple dipoles. First of all, these non-local equations extend the local variational equations derived by Netz and Orland~\cite{netzvar} beyond the dielectric continuum limit. Secondly,  the SC relations embody electrostatic correlation effects neglected in our previous work in Ref.~\cite{NLPB1}. The  non-local SC equations are solved in Section~\ref{dia} at one-loop order in order to understand the effect of solvent confinement on the electrostatic energy barrier for ionic penetration into nanopores. Then, within a restricted variational approach, Section~\ref{dia} deals with the dielectric anisotropy and reduction effects observed in solvents under nanoconfinement~\cite{lang,hansim}. Finally, in Section~\ref{cap}, we compare the prediction of the present formalism with experimental capacitance data in order to elucidate the importance of solvent structure and local/non-local correlation effects. Our results, as well as the limitations of the theory and possible improvements are discussed in the Conclusions.

\section{General self-consistent formalism}
\label{genvar}
\subsection{Dipolar liquid model and non-local self-consistent equations}

\begin{figure}
\includegraphics[width=1.15\linewidth]{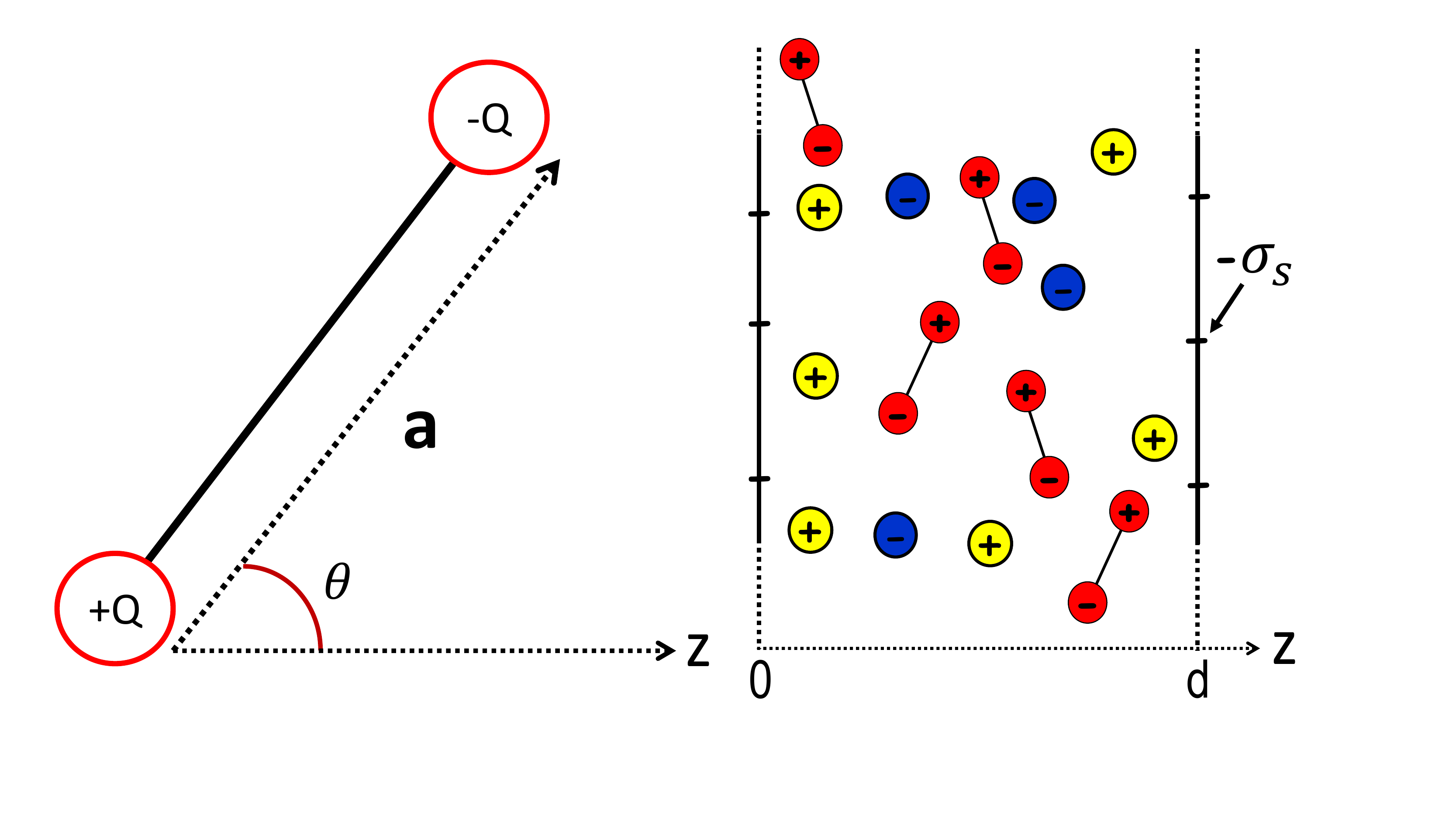}
\caption{(Color online)  Left : Charge geometry of the solvent molecules with size $a=1$ {\AA}; the charges of valency $Q=1$ are placed at the ends. Right : Geometry of the slit nanopore with surface charge $-\sigma_s\leq 0$ 
confining solvent molecules (red dipoles), anions (blue circles), and cations (yellow circles).}
\label{fig1}
\end{figure}

In this section, we reintroduce the polar liquid model of Ref.~\cite{NLPB1} and then derive the corresponding SC equations that will allow to consider charge correlation effects in confined solvents. The solvent charge structure and the composition of the charged fluid in the nanoslit are depicted in Fig.~\ref{fig1}. The liquid is composed of an arbitrary number of ionic species $i=1...p$ with each species of valency $q_i$. Ions are immersed in a solvent composed of finite size dipoles, each dipole corresponding to two elementary charges of opposite sign $\pm Q$ separated by the fixed distance $a$.  In Ref.~\cite{NLPB1}, the grand-canonical partition function of the liquid in the form of a functional integral over a fluctuating electrostatic potential $ Z_G=\int \mathcal{D}\phi\;e^{-H[\phi]}$ has been derived, with the Hamiltonian functional
\bea\label{HamFunc}
H[\phi]&=&\frac{k_BT}{2e^2}\int\mathrm{d}\br\;\e_0(\br)\left[\nabla\phi(\br)\right]^2-i\int\mathrm{d}\br\sigma(\br)\phi(\br)\nonumber\\
&&-\Lambda_s\int\frac{\mathrm{d}\br\mathrm{d}\bom}{4\pi}e^{E_s-W_s(\br,\ba)}e^{iQ\left[\phi(\br)-\phi(\br+\ba)\right]}\nonumber\\
&&-\sum_i\Lambda_i\int\mathrm{d}\br e^{E_i-W_i(\br)}e^{iq\phi(\br)},
\eea
where $k_BT$ is the thermal energy, $e$ the elementary charge, and the function $\e_0(\br)$ accounts for the dielectric permittivity difference between vacuum and the membrane of permittivities $\e_0$ and $\e_m$, respectively. We note that in the present work, the dielectric permittivities will be given in units of the vacuum permittivity, which is equivalent to setting $\e_0=1$. However, for the sake of generality, the coefficient $\e_0$ will be kept in the equations. Furthermore, $\sigma(\br)$ denotes the fixed charge distribution on the pore walls, and $\Lambda_i$ and $\Lambda_s$ stand for the fugacity of ions and solvent molecules. 

In Eq.~(\ref{HamFunc}), the first term is the electrostatic energy of freely propagating waves in vacuum, the second term couples the surface charge to the fluctuating potential, and the third and fourth terms correspond to the number density of mobile ions and solvent molecules, respectively. Then, the self energy of ions and polar molecules in vacuum are given by
\bea\label{self1}
E_i&=&\frac{q_i^2}{2}v_c(0)\\
\label{self2}
E_s&=&Q^2v_c(0).
\eea
In Eqs.~(\ref{self1}) and~(\ref{self2}), the bulk Coulomb potential in vacuum is $v_c(r)=\ell_B/r$, with $\ell_B=e^2/(4\pi\e_0k_BT)\simeq 55$ nm being the Bjerrum length. Moreover, the general wall potentials $W_i(\br)$ and $W_s(\br,\ba)$  for ions and solvent molecules account for the presence of any impenetrable boundaries in the system. We also note in passing that by taking the point-dipole limit of Eq.~(\ref{HamFunc}), which consists in Taylor-expanding the dipolar term at the quadratic order in the solvent molecular size $a$, one obtains the Hamiltonian of the point-dipole liquid introduced in Ref.~\cite{dunyuk}. Starting with this point-dipole model, in Ref.~\cite{epl} an extended dipolar PB formalism (EDPB) incorporating interfacial correlation effects on the solvent dielectric response has been developed. In Section~\ref{cap}, we will reconsider the local EDPB formalism for comparison with the more general non-local model of the present work. 

The polar liquid model of Eq.~(\ref{HamFunc}) was considered in Ref.~\cite{NLPB1} at the MF level. In order to analyze non-local correlation effects beyond MF theory, we now derive the corresponding self-consistent equations. The first step consists in computing the variational grand potential 
\be\label{var1}
\Omega_v=\Omega_0+\lan H-H_0\ran_0,
\ee
with the reference Hamiltonian of the most general quadratic form
\be\label{ref}
H_0\left[\phi\right]=\int\frac{\mathrm{d}\br\mathrm{d}\br'}{2}\left[\phi-i\phi_0\right]_\br v_0^{-1}(\br,\br')\left[\phi-i\phi_0\right]_{\br'},
\ee
where the trial functions are the external electrostatic potential $\phi_0(\br)$ and the propagator $v_0(\br,\br')$. In Eq.~(\ref{var1}), the van der Waals contribution resulting from quadratic fluctuations reads as
\be\label{vdwcon}
\Omega_0=-\ln\int\mathcal{D}\phi\;e^{-H_0\left[\phi\right]}=-\frac{1}{2}\mathrm{Tr}\ln\left[v_0\right],
\ee
and the field-theoretic average of a general functional $F[\phi]$ with respect to the reference Hamiltonian is defined as
\be
\lan F\left[\phi\right]\ran_0\equiv\frac{\int\mathcal{D}\phi\;e^{-H_0\left[\phi\right]}F\left[\phi\right]}
{\int\mathcal{D}\phi\;e^{-H_0\left[\phi\right]}}.
\ee
Evaluating the field-theoretic averages in Eq.~(\ref{var1}), the variational grand potential follows in the form
\begin{widetext}
\bea\label{varf}
\Omega_v&=&-\frac{1}{2}\mathrm{Tr}\ln\left[v_0\right]+\int\mathrm{d}\br\sigma(\br)\phi_0(\br)+\frac{k_BT}{2e^2}\int\mathrm{d}\br\left\{\e_0(\br)\nabla_\br\cdot\nabla_{\br'} \left.v_0(\br,\br')\right|_{\br'\to\br}-\e_0(\br)\left[\nabla\phi_0(\br)\right]^2\right\}\\
&&-\sum_i\Lambda_i\int\mathrm{d}\br e^{E_i-W_i(\br)}e^{-q_i\phi_0(\br)}e^{-\frac{q_i^2}{2}v_0(\br,\br)}-\Lambda_s\int\mathrm{d}\br\frac{d\bom}{4\pi}e^{E_s-W_s(\br,\ba)}e^{-Q\left[\phi_0(\br)-\phi_0(\br+\ba)\right]}\; e^{-\frac{Q^2}{2}v_d(\br,\ba)}.\nonumber
\eea
\end{widetext}
We note that the forth and fifth terms on the r.h.s. of the grand potential~(\ref{varf}) correspond, respectively, to the average density of ions and dipoles. We also introduced in the fifth term the dipolar self-energy defined as
\be\label{dipself}
v_d(\br,\ba)=v_0(\br,\br)+v_0(\br+\ba,\br+\ba)-v_0(\br,\br+\ba)-v_0(\br+\ba,\br).
\ee

The ionic number density is determined from the grand potential according to the relation $\rho_i(\br)=\delta\Omega_v/\delta W_i(\br)$, which yields
\be\label{ion1}
\rho_i(\br)=\Lambda_i e^{E_i-W_i(\br)}e^{-q_i\phi_0(\br)}e^{-\frac{q_i^2}{2}v_0(\br,\br)}.
\ee
From the bulk limit of Eq.~(\ref{ion1}), the relation between the ionic fugacity and the reservoir concentration is obtained in the form
\be\label{bi}
\rho_{ib}=\Lambda_i\exp\left[E_i-\frac{q_i^2}{2}v_0^b(0)\right],
\ee
where $v_0^b(0)$ is the ionic self energy in a bulk solvent, i.e. the equal-point electrostatic propagator in the absence of any boundaries. In order to obtain the number density of the two elementary charges located at the ends of the solvent molecule, we split the wall potential into two parts, $W_s(\br,\ba)=W_+(\br)+W_-(\br+\ba)$, where the functions $W_+(\br)$ and $W_-(\br+\ba)$ respectively are the steric potentials experienced by the negative and positive charges on the solvent molecule (the origin of the molecule located at $\br$ corresponds to the positive charge, see Fig.~\ref{fig1}). By taking the functional derivatives of the grand potential~(\ref{varf}) with respect to the potential $W_\pm(\br,\ba)$, the number densities for the solvent follow in the form
\be\label{densol}
\rho_{s\pm}(\br)=\int\frac{d\bom}{4\pi} f_{s\pm}(\br,\ba),
\ee
with the solvent densities at the fixed orientation $\bom$ defined as
\bea\label{denomsp}
f_{s+}(\br,\ba)&=&\Lambda_se^{E_s-W_s(\br,\ba)}e^{-Q\left[\phi_0(\br)-\phi_0(\br+\ba)\right]}\\
&&\times e^{-\frac{Q^2}{2}v_d(\br,\ba)}\nonumber\\
\label{denomsm}
f_{s-}(\br,\ba)&=&\Lambda_se^{E_s-W_s(\br-\ba,\ba)}e^{-Q\left[\phi_0(\br-\ba)-\phi_0(\br)\right]}\\
&&\times e^{-\frac{Q^2}{2}v_d(\br-\ba,\ba)}.\nonumber
\eea
One also notes that these functions are related to each other according to the relation $f_{s-}(\br,\ba)=f_{s+}(\br-\ba,\ba)$, as expected. Finally, from the bulk limit of Eq.~(\ref{densol}), one gets the relation between the solvent fugacity and the reservoir density in the form
\be\label{bs}
\rho_{sb}=\Lambda_s\exp\left\{E_s-Q^2\left[v_0^b(0)-v_0^b(a)\right]\right\}.
\ee

By taking the functional derivative of the grand potential~(\ref{varf}) with respect to the trial potential $\phi_0(\br)$, the equation of state determining the external potential follows as
\begin{widetext}
\bea\label{varnlpb1}
\frac{k_BT}{e^2}\nabla_\br\e_0(\br)\nabla_\br\phi_0(\br)+\sum_iq_i\rho_{i}(\br)+Q\left[\rho_{s+}(\br)-\rho_{s-}(\br)\right]=-\sigma(\br).
\eea
Then, the functional derivative of the grand potential~(\ref{varf}) with respect to the propagator $v_0(\br,\br')$ results in the following relation for the electrostatic kernel,
\bea\label{kernel}
v_0^{-1}(\br,\br')&=&-\frac{k_BT}{e^2}\nabla_\br\e_0(\br)\nabla_\br\delta(\br-\br')+\sum_iq_i^2\rho_i(\br)\delta(\br-\br')\nonumber\\
&&+Q^2\int\frac{\mathrm{d}\bom}{4\pi}\left\{f_{s+}(\br,\ba)\left[\delta(\br'-\br)-\delta(\br'-\br-\ba)\right]+f_{s-}(\br,\ba)\left[\delta(\br'-\br)-\delta(\br'-\br+\ba)\right]\right\}.
\eea
Using the definition of the Green's function
\be\label{defgr}
\int\mathrm{d}\br''\;v_0^{-1}(\br,\br'')v_0(\br'',\br')=\delta(\br-\br'),
\ee
one can invert the kernel~(\ref{kernel}) and finally obtain the variational equation for the corresponding electrostatic Green's function as
\bea\label{var2}
&&-\frac{k_BT}{e^2}\nabla_\br\e_0(\br)\nabla_\br v_0(\br,\br')+\sum_iq_i^2\rho_i(\br)v_0(\br,\br')\nonumber\\
&&+Q^2\int\frac{\mathrm{d}\bom}{4\pi}\left\{f_{s+}(\br,\ba)\left[v_0(\br,\br')-v_0(\br+\ba,\br')\right]+f_{s-}(\br,\ba)\left[v_0(\br,\br')-v_0(\br-\ba,\br')\right]\right\}=\delta(\br-\br').
\eea
\end{widetext}

The equations~(\ref{varnlpb1}) and~(\ref{var2}) that incorporate the solvent charge structure generalize the solvent-implicit variational equations by Netz and Orland of Ref.~\cite{netzvar}. The first integro-differential equation~(\ref{varnlpb1}) is a non-local PB (NLPB) equation including charge fluctuation effects embodied in the ionic and dipolar self-energies (see Eqs.~(\ref{ion1}) and~(\ref{densol})-(\ref{denomsm})). In the absence of correlations, where these ionic and dipolar 
self-energies vanish, equation~(\ref{varnlpb1}) reduces to the MF NLPB equation derived in Ref.~\cite{NLPB1}. Then, the second equation~(\ref{var2}) for the electrostatic Green's function is a solvent-explicit Debye-H\"{u}ckel (DH) equation of non-local form. We emphasize that the non-local or integro-differential form of these equations stems from the extended charge structure of solvent molecules. We finally note that the equivalent of Eq.~(\ref{var2}) restricted to bulk liquids was derived in Ref.~\cite{NLPB2}, and it was shown that in the bulk limit, the associated dielectric permittivity is given by the well-known Debye-Langevin relation
\be\label{dibu}
\e_w=\e_0+\frac{4\pi}{3}\ell_BQ^2a^2\rho_{sb}.
\ee
In the present work, the solvent molecular size will be set to $a=1$ {\AA}. For solvents of physiological concentration $\rho_{sb}=55$ M, this gives the bulk permittivity value $\e_w=76.75\;\e_0$. 

\subsection{Non-local self-consistent equations in slit nanopores}

Since the present work considers confinement effects in slit geometries, in this part, we can simplify the general equations~(\ref{varnlpb1}) and~(\ref{var2}) for a planar geometry. The polar liquid and ions are confined to a 
slit nanopore with rigid interfaces located at $z=0$ and $z=d$ (see Fig.~\ref{fig1}). In the slit geometry, the dielectric permittivity function is given by
\be
\label{dislit}
\e_0(\br)=\e_0(z)=\e_0\theta(z)\theta(d-z)+\e_m\theta(-z)\theta(z-d),
\ee
with the vacuum permittivity $\e_0=1$ and the membrane permittivity $\e_m$. The ionic confinement is imposed by the wall potential $W_i(\br)=W_i(z)=0$ if $0\leq z\leq d$ and $W_i(z)=\infty$ otherwise. 
Moreover, in terms of the projection of the dipolar alignment on the $z$ axis 
\be\label{az}
a_z=a\cos\theta,
\ee
where $\theta$ is the angle between the dipole and the $z$ axis (see Fig.~\ref{fig1}), the dipolar wall potential imposing the solvent confinement is given by $W_s(\br,\ba)=W_s(z,a_z)=0$ if $0\leq z\leq d$ and $0\leq z+a_z\leq d$, and $W_s(z,a_z)=\infty$ otherwise.

Exploiting now the translational symmetry in the $(x,y)$-plane within the planar geometry, one can Fourier-expand the Green's function as
\be\label{four1}
v_0(\br,\br')=\int\frac{\mathrm{d}^2\bk}{4\pi^2}e^{i\bk\cdot\left(\br_\pa-\br'_\pa\right)}\tv_0(z,z';k).
\ee
To simplify the notation from now on, we will omit the $k$-dependence of the Fourier-transformed Green's function. Carrying out in Eq.~(\ref{four1}) the integral over the angle $\theta_\bk$ in the reciprocal plane, one obtains
\be\label{four2}
v_0(\br,\br')=\int_0^\Lambda\frac{\mathrm{d}kk}{2\pi}J_0\left[k|\br_\pa-\br'_\pa|\right]\tv_0(z,z'),
\ee
with the ultraviolet (UV) cut-off $\Lambda$ and the Bessel function of the first kind $J_0(x)$. Injecting the expansion in Eq.~(\ref{four2}) into Eq.~(\ref{dipself}), the dipolar self-energy follows in the form
\bea\label{dipself2}
v_d(z,a_z)&=&\int_0^\Lambda\frac{\mathrm{d}kk}{2\pi}\left\{\tv_0(z,z)+\tv_0(z+a_z,z+a_z)\right.\nonumber\\
&&\hspace{1.7cm}-2\tv_0(z,z+a_z)\left.J_0\left[k|a_\pa|\right]\right\}\nonumber,\\
\eea
where we defined the projection of the dipolar vector $\ba$ onto the $(x,y)$-plane
\be\label{ap}
a_\pa=a\sin\theta.
\ee
We note that  the amplitude of the parallel component $a_\pa$ in Eq.~(\ref{dipself2}) is related to the perpendicular component $a_z$ through the relation
\be
|a_\pa|=\sqrt{a^2-a_z^2}.
\ee
Using now in Eqs.~(\ref{ion1}) and~(\ref{densol})-(\ref{denomsm}) the bulk relations~(\ref{bi}) and~(\ref{bs}) between the particle densities and fugacities, and defining the renormalized ionic and dipolar self-energies
\bea
\label{is2i}
&&\delta v_i(z)=v_0(z,z)-v_0^b(0)\\
\label{is2d}
&&\delta v_d(z,a_z)=v_d(z,a_z)-2v_0^b(0)+2v_0^b(a),
\eea
the ionic and solvent number densities follow in the form
\bea
\label{id}
&&\rho_i(z)=\rho_{ib}e^{-q_i\phi_0(z)}e^{-\frac{q_i^2}{2}\delta v_i(z)-W_i(z)}\\
\label{sd}
&&\rho_{s\pm}(z)=\int_{a_1(z)}^{a_2(z)}\frac{\mathrm{d}a_z}{2a}f_{s\pm}(z,a_z),
\eea
with the solvent molecular charged densities at fixed orientation
\be\label{denan}
f_{s\pm}(z,a_z)=\rho_{sb}e^{-\frac{Q^2}{2}\delta v_d(z,a_z)}e^{\pm Q\left[\phi_0(z+a_z)-\phi_0(z)\right]}.
\ee
In Eq.~(\ref{sd}), we introduced the integral boundaries taking into account the impenetrability of the interfaces,
\bea
a_1(z)&=&-\mathrm{min}(a,z)\\
a_2(z)&=&\mathrm{min}(a,d-z).
\eea
Furthermore, in passing from Eq.~(\ref{densol}) to Eq.~(\ref{sd}), we performed the change of variable $\theta\to a_z$ in the integral over the dipole rotations. Substituting now the solvent density~(\ref{sd}) into the variational NLPB Eq.~(\ref{varnlpb1}), and inserting the Fourier expansion of the Green's function~(\ref{four1}) and the solvent density function~(\ref{denan}) into the second variational equation~(\ref{var2}),  after some lengthy algebra, the electrostatic self-consistent equations take for $0\leq z\leq d$ the simpler form
\begin{widetext}
\bea\label{varnlpb2}
&&\frac{k_BT}{e^2}\partial_z\e_0(z)\partial_z\phi_0(z)+\sum_iq_i\rho_i(z)+2Q\rho_{sb}\int_{a_1(z)}^{a_2(z)}\frac{\mathrm{d}a_z}{2a}\sinh\left[Q\phi_0(z+a_z)-Q\phi_0(z)\right]\;e^{-\frac{Q^2}{2}\delta v_d(z,a_z)}=-\sigma(z)\\
\label{var3}
&&-\frac{k_BT}{e^2}\left[\partial_z\e_0(z)\partial_z-\e_0(z)p^2(z)\right]\tv_0(z,z')\\
&&+2Q^2\rho_{sb}\int_{a_1(z)}^{a_2(z)}\frac{\mathrm{d}a_z}{2a}\cosh\left[Q\phi_0(z+a_z)-Q\phi_0(z)\right]\; e^{-\frac{Q^2}{2}\delta v_d(z,a_z)}\left\{\tv_0(z,z')-\tv_0(z+a_z,z')J_0(k|a_\pa|)\right\}=\delta(z-z'),\nonumber
\eea
\end{widetext}
In Eq.~(\ref{var3}), we introduced the auxiliary function
\be
p(z)=\sqrt{k^2+\kappa_i^2(z)}.
\ee
with the ionic screening function
\be
\kappa_i^2(z)=\frac{e^2}{\e_0(z) k_BT}\sum_iq_i^2\rho_i(z).
\ee

This concludes the derivation of the final form of the equations which we will solve in the following in different settings. 

%In Section~\ref{dilsol} we will consider a dilute solvent confined to a neutral slit characterised by a vanishing external potential $\phi_0(\br)=0$. 
%In this case we will solve Eq.~(\ref{var3}) at the one-loop level within an iterative scheme. Because this iterative scheme fails for solvents at physiological concentrations, the latter case will be considered in Section~\ref{dia} within a restricted %variational scheme in order to investigate dielectric reduction and anisotropy effects in nanopores. Finally, in Section~\ref{cap}, the correlation corrected NLPB equation~(\ref{varnlpb2}) will be solved by using a relaxation algorithm in order to %scrutinize non-local correlation effects on the dielectric response of the solvent at charged interfaces and the differential capacitance of materials with low polarity.

\section{Results}
\subsection{Dilute solvents in slit nanopores}
\label{dilsol}

As a first application of the theory we consider in this part the polar liquid confined to a neutral slit $\sigma_s=0$, and containing a dilute symmetric electrolyte composed of monovalent anions and cations. 
According to Eq.~(\ref{varnlpb2}), this corresponds to a vanishing external potential $\phi_0(z)=0$. To characterize electrostatic correlations in the nano-slit, the remaining variational equation~(\ref{var3}) will be 
considered at the one-loop level. The one-loop approximation consists in linearizing the equation in terms of the propagator, or equivalently neglecting the dipolar self-energies in the exponentials. This yields a 
dipolar DH-equation in the form
\begin{widetext}
\be\label{lap}
-\frac{k_BT}{e^2}\left[\partial_z\e_0(z)\partial_z-\e_0(z)p^2_b\right]\tv_0(z,z')+2Q^2\rho_{sb}\int_{a_1(z)}^{a_2(z)}\frac{\mathrm{d}a_z}{2a}\left\{\tv_0(z,z')-\tv_0(z+a_z,z')J_0(k|a_\pa|)\right\}=\delta(z-z'),
\ee
\end{widetext}
with the coefficients
\bea
p_b&=&\sqrt{k^2+\kappa_{ib}^2}\\
\kappa_{ib}^2&=&\frac{e^2}{\e_0 k_BT}\sum_iq_i^2\rho_{ib}\theta(z)\theta(d-z).
\eea
\begin{figure*}
\includegraphics[width=.45\linewidth]{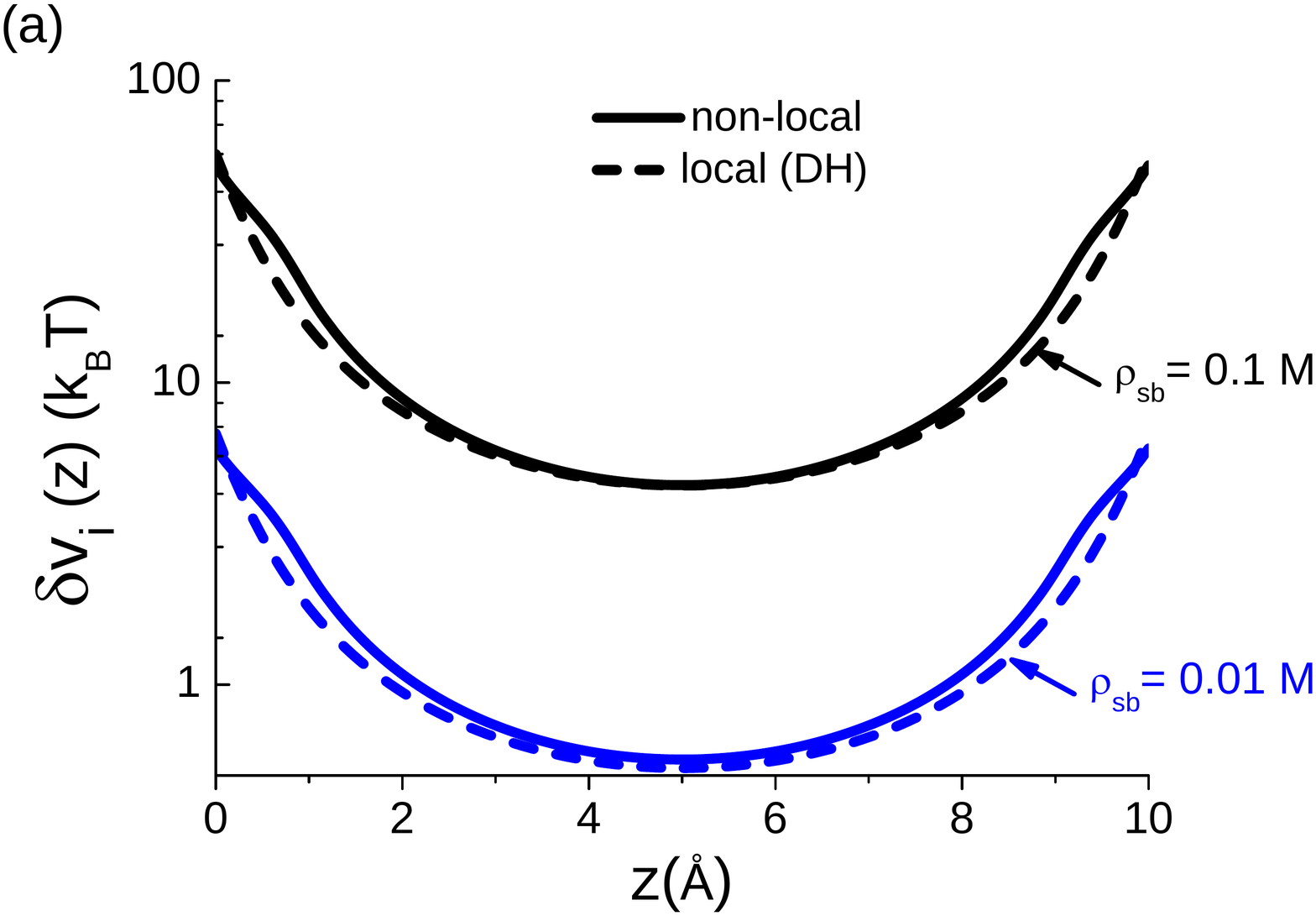}
\includegraphics[width=.45\linewidth]{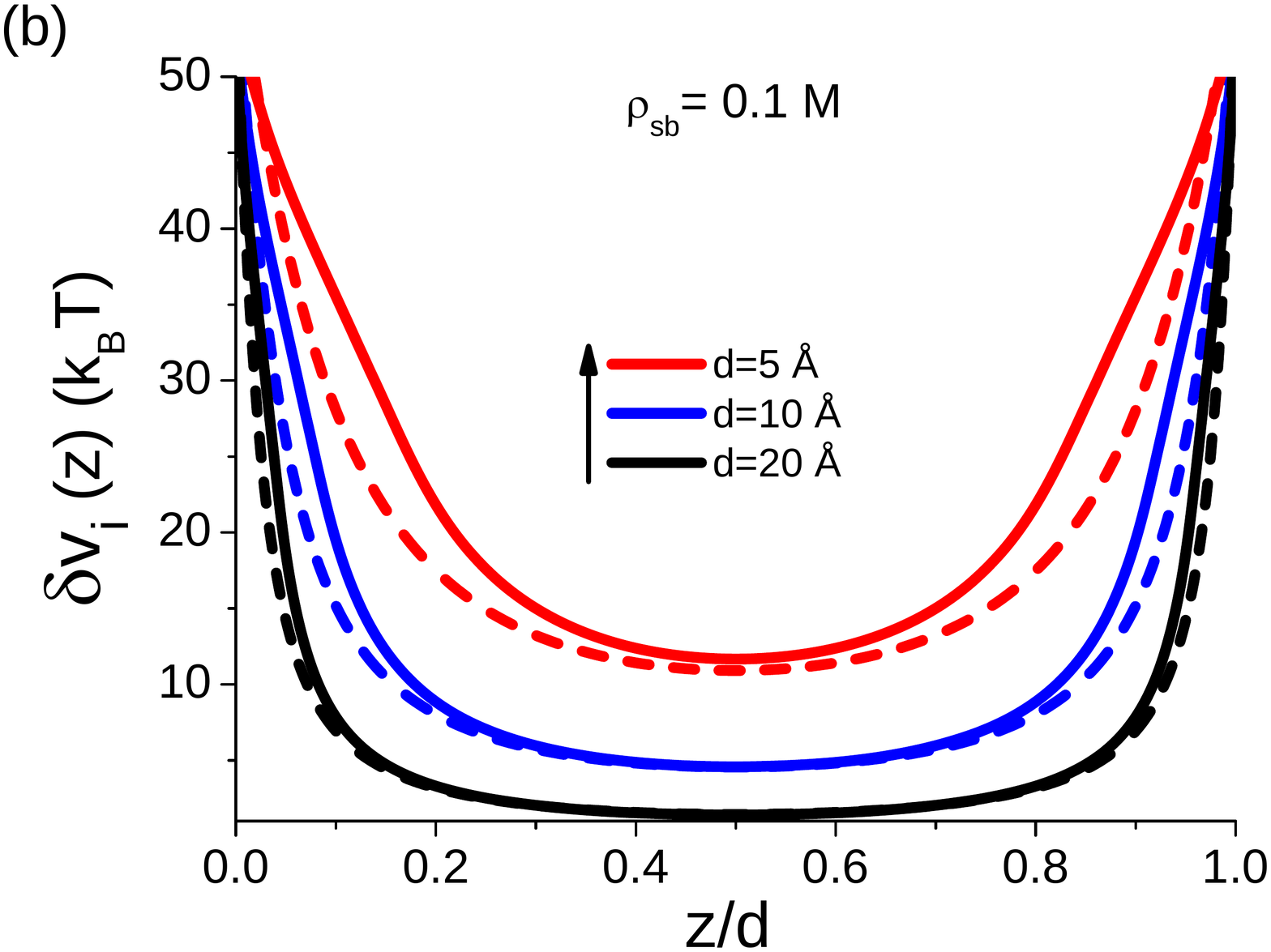}
\includegraphics[width=.45\linewidth]{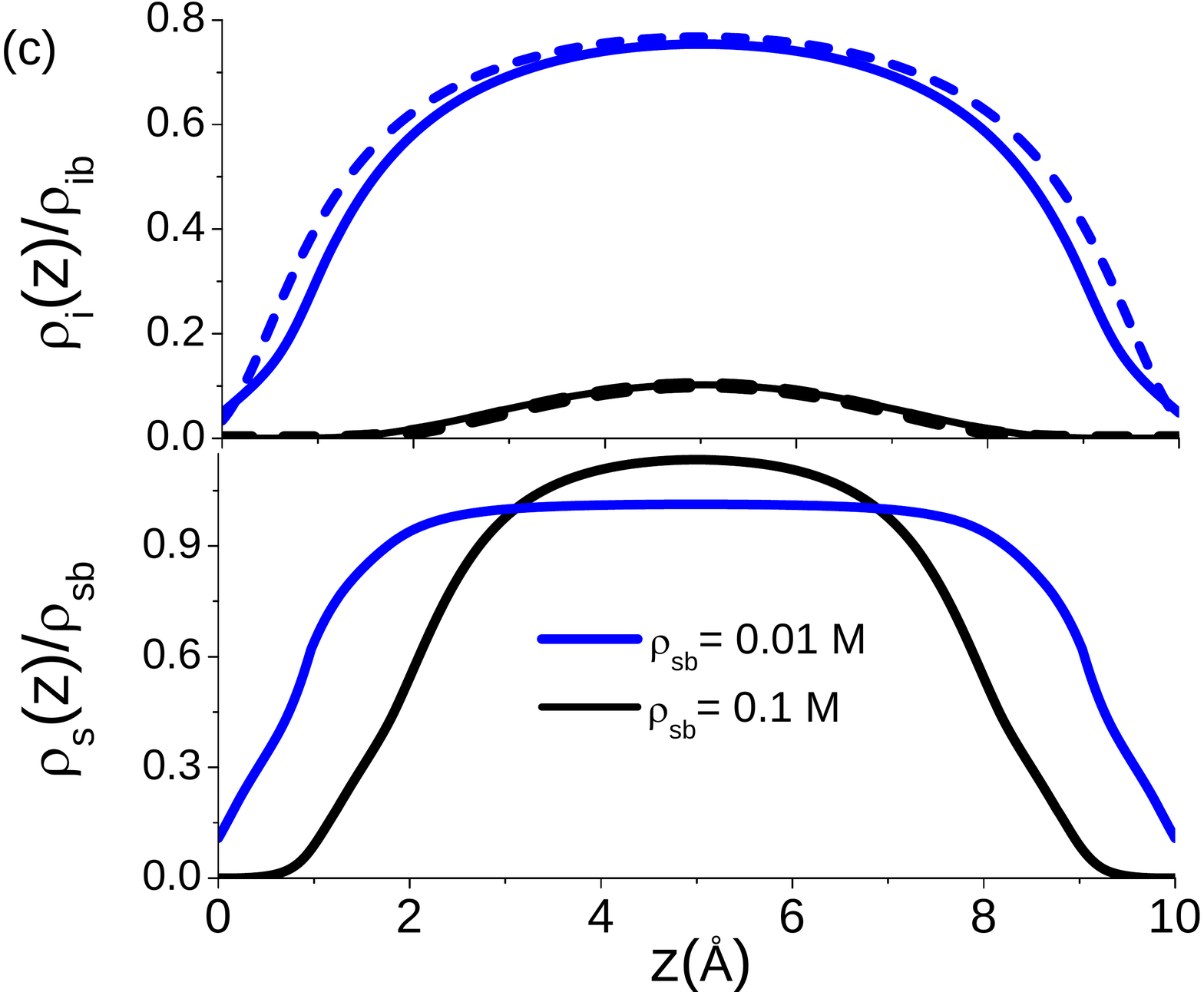}
\includegraphics[width=.45\linewidth]{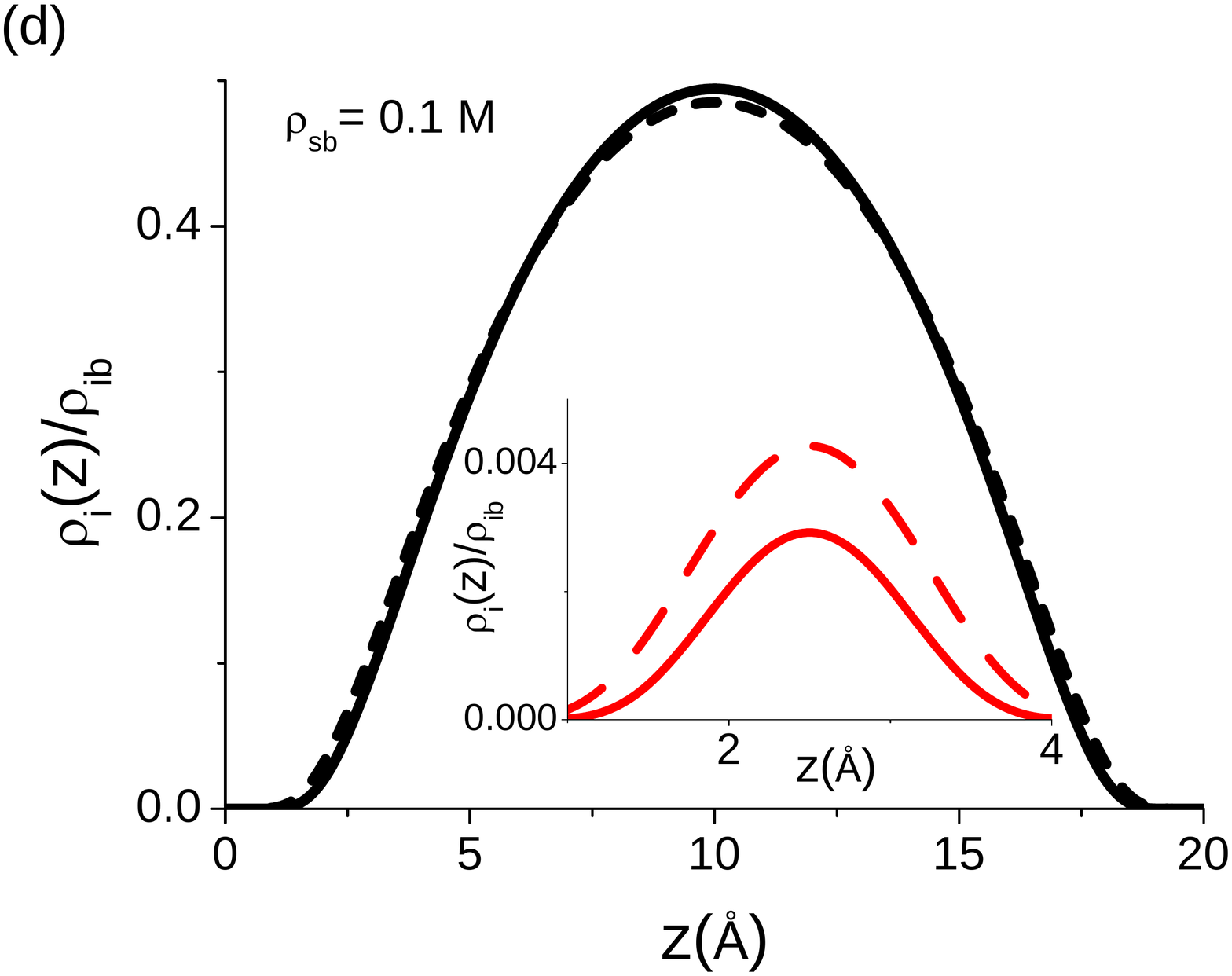}
\caption{(Color online)  Ionic self-energies for (a) two values of the bulk solvent density at the pore size $d=10$ {\AA} and (b) different pore sizes at the solvent concentration $\rho_{sb}=0.1$ M.  (c) Local ion (top) and solvent densities (bottom) in a slit-pore of size $d=10$ {\AA}. The solvent densities in the reservoir are $\rho_{sb}=0.01$ M (blue curves) and $0.1$ M (black curves). (d) Ion densities for pores of size $d=5$ {\AA} (inset) and $d=20$ {\AA} (main plot) at the reservoir density $\rho_{sb}=0.1$ M. In all figures, solid and dashed curves correspond, respectively, to the solutions obtained from the non-local Eq.~(\ref{lap}) and the dielectric continuum Eq.~(\ref{dielcon}). The bulk ion concentration is $\rho_{ib}=10^{-6}$ M, and the slit is neutral ($\sigma_s=0.0$ e $\mbox{nm}^{-2}$).}
\label{fig2}
\end{figure*}

At this stage, it is instructive to note that the DH-equation of the dielectric continuum description follows from the point-dipole limit of Eq.~(\ref{lap}). More precisely, by expanding Eq.~(\ref{lap}) up to quadratic order in the solvent molecular 
size $a$, relaxing the rotational penalty for dipoles by setting $a_1(z)=-a$ and $a_2(z)=a$, and carrying out the integral over the dipole rotations, one ends up with the usual Fourier-expanded DH-equation in planar geometry,
\be\label{dielcon}
\left\{\partial_z\e(z)\partial_z-\e(z)\left(k^2+\kappa_{DH}^2\right)\right\}\tv_{DH}(z,z')=-\frac{e^2}{k_BT}\delta(z-z'),
\ee
with the dielectric permittivity function 
\be
\e(z)=\e_0\left[\theta(-z)+\theta(z-d)\right]+\e_w\theta(z)\theta(d-z)
\ee
 and the DH screening parameter
\be\label{dh}
\kappa_{DH}^2=\frac{e^2}{\e_wk_BT}\sum_iq_i^2\rho_{ib}.
\ee

We solved Eq.~(\ref{lap}) by using a numerical inversion scheme detailed in Appendix~\ref{ap1}, with the bulk salt density set to $\rho_{ib}=10^{-6}$ M. We note in passing that because our solvent model is made up of finite-size dipoles, Eqs.~(\ref{varnlpb2})-(\ref{var3}) and Eq.~(\ref{lap}) do not present any UV-divergence. However, in order to simplify the tremendous numerical task, the integrals in Fourier space were computed with a finite ultraviolet (UV) cut-off $\Lambda=1000/\ell_B$.  The ionic self-energies defined in Eq.~(\ref{is2i}) are reported in Fig.~\ref{fig2}(a) for dilute solvents with density $\rho_{sb}=0.01$ M (solid blue curves) and $0.1$ M (solid black curves). We also display as dashed curves the local image charge potentials corresponding to the solution of Eq.~(\ref{dielcon}) with the finite cut-off $\Lambda=1000/\ell_B$ (see the end of Appendix~\ref{ap1} for the explicit form of the DH image-charge potential).  First of all, one sees that for both solvent concentrations, the ionic self-energies in the dipolar and dielectric continuum liquids are very close to each other, the non-local potential being only slightly higher than the DH potential close to the pore walls. This remarkable result shows that the explicit consideration of solvent interactions at the one-loop level naturally results in the ``image charge'' forces usually obtained by imposing the dielectric jump between the membrane medium and the solvent. 

In Fig.~\ref{fig2}(a), one also notes that the rise of the solvent density from $\rho_{sb}=0.01$ M to 0.1 M that amplifies the dielectric screening ability of the pore with respect to the membrane increases the amplitude of the ionic self-energy by almost an order of magnitude. In the top plot of Fig.~\ref{fig2}(c), we see that this results in a stronger ionic exclusion from the pore.  More interestingly, in the bottom plot of Fig.~\ref{fig2}(c), the same effect is shown to act as a hydrophobic force on solvent molecules, shifting their position far away from the interface. \textcolor{black}{Thus, the reduced dielectric permittivity of the membrane directly results in the solvophobicity of the interface.} We also considered the effect of the confinement on the ionic self-energies and densities. Fig.~\ref{fig2}(b) shows that the reduction of the pore size is qualitatively equivalent to the increase of the bulk solvent concentration, that is, it amplifies the energetic barrier for charge penetration. In Fig.~\ref{fig2}(d), this is shown  to result in a stronger ion rejection (compare the main plot and the inset). More importantly, in Fig.~\ref{fig2}(b), one notes that the difference between the local and the non-local self-energy becomes relevant for pores of subnanometer size, i.e. if the confinement scale becomes comparable to the size of solvent molecules. 

Within the present solvent-explicit theory, the energetic barrier for ionic penetration into the pore is composed of two contributions. First, the pore dielectric permittivity being larger than the membrane permittivity, mobile charges experience a stronger dielectric screening in the  mid-pore region than close to the membrane. Hence, ions feel a repulsive force excluding them from the pore wall, which is the well-known ``image-charge'' effect already present in the dielectric continuum electrostatics. The second contribution is the Born energy difference between the pore and the bulk reservoir. Due to the confinement in the pore, the solvent density and dielectric permittivity are lower than in the bulk reservoir. As a result, ions possess a lower electrostatic free energy in the reservoir and this favors their rejection from the nanoslit. This ionic Born energy absent in the dielectric continuum formulation of electrostatics is the factor increasing the non-local ionic self energies above the local results in Figs.~\ref{fig2}(a) and (b).  Before concluding, we emphasize that the one-loop level results discussed in this part neglect the effect of the dipolar self-energy $\delta v_d(z)$ present in the SC eq.~(\ref{var3}).  
In the next section where we will introduce a restricted variational scheme, the consideration of these self-energies at the SC level will be shown to result in a stronger deviation of the non-local formulation from the dielectric continuum picture.

\subsection{Solvents at physiological concentrations in slit nanopores}
\label{dia}

Deviations from the bulk behaviour in confined systems were e.g. seen in recent ion rejection experiments where it was observed that the confinement of the solvent in membrane nanopores results in a reduction of the pore dielectric 
below the bulk value~\cite{lang}. MD simulations with explicit solvent by Balleneger and Hansen found that polar liquids confined in nanoslits are characterized by a dielectric anisotropy associated with a transverse permittivity along the membrane wall $\epa$ exceeding the perpendicular component $\epe$~\cite{hansim}, a feature clearly absent in the dielectric continuum formulation of electrostatics. It is thus an important task to characterize the underlying physics behind these effects, and to determine the characteristic pore sizes and fluid densities where they become relevant. In order to tackle these questions we consider a salt free solvent ($\rho_{ib}=0.0$ M) confined in a neutral slit $\sigma_s=0.0$ e $\mbox{nm}^{-2}$, which results in a vanishing external potential $\phi_0(\br)=0$. Our iterative method employed in Section~\ref{dilsol} fails for high solvent densities. Instead we use a restricted self-consistent approach which is however 
general enough to capture these effects. 

\subsubsection{Local self-consistent ansatz}
\label{locsel}

Our restricted self-consistent approach consists in computing the variational grand potential~(\ref{var1}) with a {\it trial ansatz} that solves a dielectrically anisotropic Laplace equation,
\be\label{anila}
\left[\nabla_{\br_\pa}\epa(z)\nabla_{\br_\pa}+\partial_z\epe(z)\partial_z\right]v_0(\br,\br')=-\frac{e^2}{k_BT}\delta(\br-\br'),
\ee
with the dielectric permittivity functions parallel and perpendicular to the pore walls being defined as
\be\label{dielcom}
\e_{\pa,\perp}(z)=\e_{\pa,\perp}\theta(z)\theta(d-z)+\e_m\left[\theta(-z)+\theta(z-d)\right],
\ee
The solution of Eq.~(\ref{anila}) is reported in Appendix~\ref{grani}. The permittivity components $\epa$ and $\epe$ in Eq.~(\ref{dielcom}) are the variational parameters whose numerical values will be obtained from the minimization of the variational grand potential~(\ref{var1}).  Computing the latter with the reference Hamiltonian~(\ref{ref}) where the  kernel is the inverse of the solution of Eq.~(\ref{anila}), one finds that the grand potential is composed of three parts,
\be
\label{vargr1}
\Omega_v=\Omega_0+\Omega_c+\Omega_d.
\ee
We report below the explicit form of these three contributions, but leave all technical details of the analytical computations to three Appendices B-D.

The van der Waals part $\Omega_0$ defined in Eq.~(\ref{vdwcon}) is computed in Appendix~\ref{vdwder}. The result reads
\bea\label{om0ani}
\Omega_0&=&\frac{Sd\Lambda^3}{12\pi}\frac{\sqrt\epa-\sqrt\epe}{\sqrt\epe}+\frac{Sd\Lambda^3}{16\pi}\ln\frac{\epe}{\e_m}\\
&&+\frac{S\Lambda^2}{8\pi}\ln\frac{(\sqrt{\epa\epe}+\e_m)^2}{4\e_m\sqrt{\epa\epe}}\nonumber\\
&&+\frac{S}{4\pi}\int_0^\Lambda\mathrm{d}kk\ln\left(1-\Delta_\gamma^2e^{-2\gamma kd}\right),\nonumber
\eea
where $S$ is the lateral surface of the membrane, and we define dielectric anisotropy and discontinuity functions as
\bea
&&\gamma=\sqrt\frac{\epa}{\epe}.\\
&&\Delta_\gamma=\frac{\sqrt{\epe\epa}-\e_m}{\sqrt{\epe\epa}+\e_m}.
\eea
The first two terms on the r.h.s. of Eq.~(\ref{om0ani}) correspond to the electrostatic energy of a hypothetical bulk medium of volume $V_p=Sd$. Furthermore, the third term is the surface tension of two decoupled interfaces separating the membrane and the solvent media of permittivities $\e_m$ and $\sqrt{\epa\epe}$, respectively. Finally, the fourth term is the interaction energy of these interfaces located at $z=0$ and $z=d$. 

The correction term $\Omega_c$ in Eq.~(\ref{vargr1}) is given by
\bea
\Omega_c&=&S\frac{k_BT}{2e^2}\int\mathrm{d}\br\left\{\left[\e_0(z)-\epa(z)\right]\nabla_{\br_\pa}\cdot\nabla_{\br'_\pa}\right.\\
&&\hspace{2.15cm}+\left.\left[\e_0(z)-\epe(z)\right]\partial_z\partial_{z'}\right\}\left.v_0(\br,\br')\right|_{\br'\to\br}.\nonumber
\eea
Plugging into this equation the anisotropic propagator of Eq.~(\ref{dhw}), taking into account the equality $\e_m=\e_0$, and carrying out the spatial integrals over the slit width, one gets after some algebra
\bea
\Omega_c&=&\frac{Sd\Lambda^3}{48\pi}\left(2\;\frac{\e_0-\epa}{\sqrt{\epe\epa}}+\gamma\frac{\e_0-\epe}{\sqrt{\epe\epa}}\right)\\
&&+\frac{S\Lambda^2}{16\pi}\frac{\Delta_\gamma}{\epa}\left[\e_0-\epa+\gamma^2\left(\e_0-\epe\right)\right]\nonumber\\
&&+\frac{S\Delta_\gamma}{8\pi\epa}\int_0^\Lambda\frac{\mathrm{d}kke^{-2\gamma kd}}{1-\Delta_\gamma^2e^{-2\gamma kd}}\nonumber\\
&&\hspace{1.5cm}\times\left\{(\e_0-\epa)(\Delta_\gamma^2+2\Delta_\gamma\gamma kd-1)\right.\nonumber\\
&&\hspace{2cm}+\left.(\e_0-\epe)(\Delta_\gamma^2-2\Delta_\gamma\gamma kd-1)\right\}.\nonumber
\eea

Finally, the computation of the contribution from the dipole density to the grand potential~(\ref{vargr1}) is particularly involved, see Appendix~\ref{dipcon}. The final result reads as
\be\label{dipom}
\Omega_d=-S\rho_{sb}\int_0^d\mathrm{d}z\int_{a_1(z)}^{a_2(z)}\frac{\mathrm{d}a_z}{2a}e^{-\frac{Q^2}{2}\delta v_d(z,a_z)},
\ee
with the dipolar self-energy
\bea\label{dipsel}
\delta v_d(z,a_z)&=&\frac{\Lambda^3\ell_B}{6}\left\{\frac{\e_0}{\sqrt{\epa\epe}}\left(a_\pa^2+\gamma a_z^2\right)-\frac{\e_0}{\e_w}a^2\right\}\\
&&+\int_0^\Lambda\frac{\mathrm{d}kk}{2\pi}\left\{\delta \tv_0(z,z)+\delta \tv_0(z+a_z,z+a_z)\right.\nonumber\\
&&\hspace{1.8cm}\left.-2\delta \tv_0(z,z+a_z)J_0(ka_\pa)\right\},\nonumber
\eea
where the function $\delta \tv_0(z,z')$ is given by Eq.~(\ref{dhwD}) of Appendix~\ref{grani}. 
The remaining task is the numerical minimization of the grand potential $\Omega_v(\epa,\epe) $ in Eq. (\ref{vargr1}) with respect to $\epa$ and $\epe$ for which we use a 
dichotomy algorithm, with the UV cut-off set to $\Lambda=200/\ell_B$. 

\subsubsection{Dielectric reduction and anisotropy in nanoslits}
\label{nedem}

\begin{figure}
\includegraphics[width=.9\linewidth]{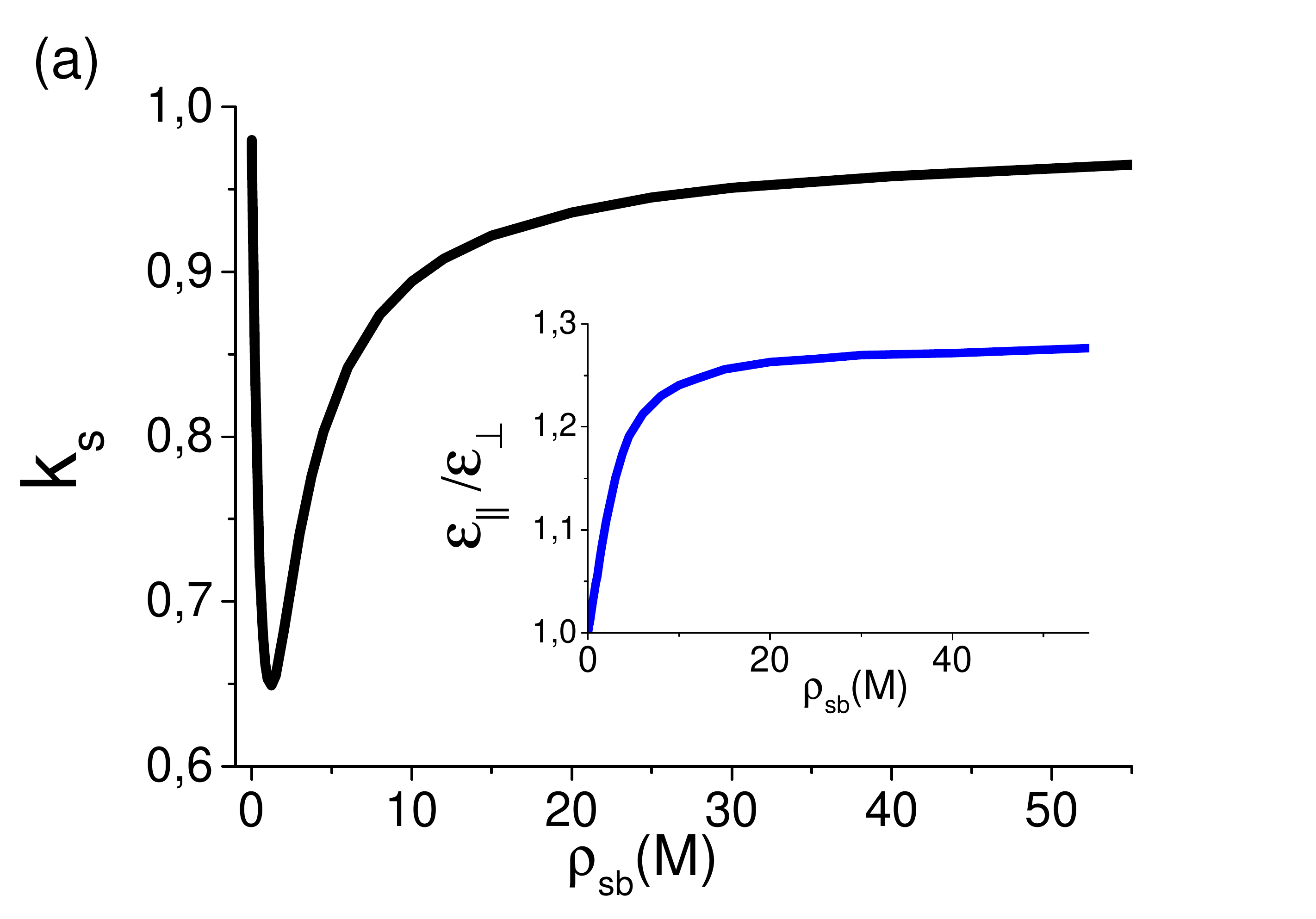}
\includegraphics[width=.9\linewidth]{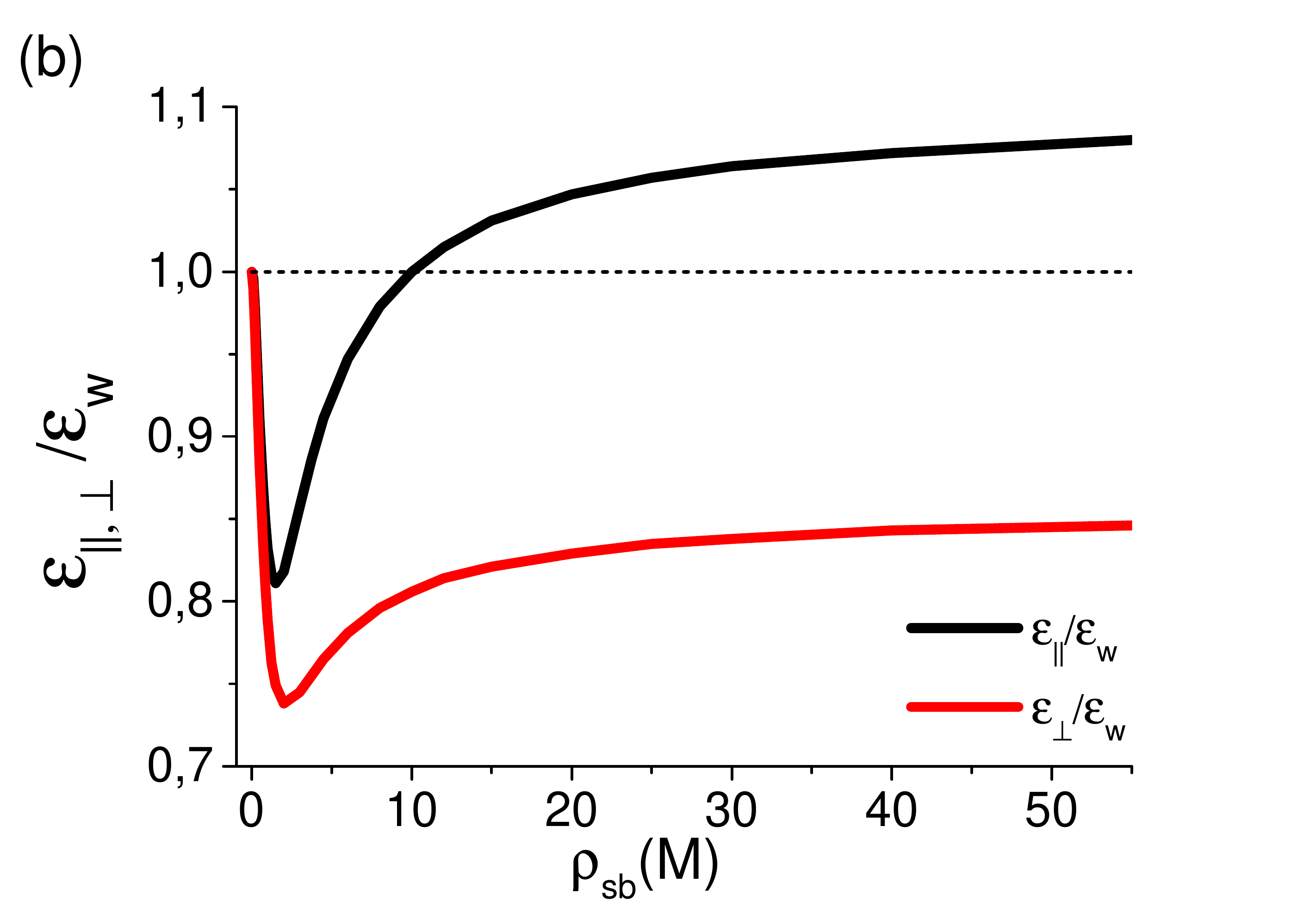}
\caption{(Color online)  (a) Solvent partition coefficient~(\ref{sopa}) and (b) dielectric permittivity components renormalized by the reservoir permittivity against the bulk solvent concentration. The inset in (a) displays the ratio of the parallel and longitudinal dielectric permittivities.The slit thickness is $d=1$ nm, and the liquid is ion-free ($\rho_{ib}=0.0$ M).}
\label{fig22}
\end{figure}

In order to characterize the dielectric reduction we use the partition coefficient defined as the pore-averaged solvent density  
\be\label{sopa}
k_s=\frac{1}{d\rho_{sb}}\int_0^d\mathrm{d}z\;\rho_{s+}(z)
\ee
(see Eq. (\ref{sd})) as a function of the bulk solvent concentration for a slit of size $d=1$ nm, plotted in Fig.~\ref{fig22}(a), which exhibits a non-monotonic behaviour. It first decreases with the bulk density for very dilute solvents, 
goes through a minimum and rises again towards $k_s=1$. Furthermore, Fig.~\ref{fig22}(b) shows that the dielectric permittivity components qualitatively follow the same trend as a function of $\rho_{sb}$. This stems from the fact that similar to a bulk liquid  (see the Debye-Langevin relation~(\ref{dibu})), the pore permittivities are increasing functions of the pore solvent density.

The non-monotonic behaviour of the quantities shown in Figs.~\ref{fig22}(a) and (b)  can be explained in terms of a competition between image dipole interactions and their dielectric screening. According to Eqs.~(\ref{sd})-(\ref{denan}), the pore solvent density is fixed by the dipolar self-energy~(\ref{dipsel}) obtained from the electrostatic propagator~(\ref{dhwD}). Eq.~(\ref{dhwD})  indicates that the amplitude of the propagator is in turn set by the coefficient $\Delta_\gamma/\sqrt{\epa\epe}\sim\Delta_0/\e_w$ where the permittivity $\e_w$ in the denominator determines the intensity of the dielectric screening of image  interactions and
\be\label{diju}
\Delta_0=\frac{\e_w-1}{\e_w+1}
\ee
is the dielectric jump function.  The function $\Delta_0/\e_w$ has its minimum at $\e_w=1+\sqrt{2}$. According to Eq.~(\ref{dibu}), this corresponds to the bulk density $\rho^*_{sb}\simeq 1$ M. In the density regime $0\leq\rho_{sb}\leq\rho^*_{sb}$, the coefficient $\Delta_0/\e_w$ is amplified with an increase of the solvent density $\rho_{sb}$, because the function $\Delta_0$ rises faster than the permittivity $\e_w$. Thus, image forces dominate in this regime the dielectric screening, and the solvent partition coefficient drops with increasing $\rho_{sb}$. Then, increasing the density $\rho_{sb}$ above the characteristic value $\rho^*_{sb}$, the dielectric jump function $\Delta_0$ saturates to one but the permittivity $\e_w$ continues to rise. This results in turn in a net reduction of image dipole forces by dielectric screening, and the partition coefficient increases towards the saturation value  $k_s=1$. These results indicate that for solvents of concentration $\rho_{sb}=55$ M confined in nanoscale pores, the strong dielectric screening of image dipole interactions result in a rather weak solvent exclusion from the pore.

The second point to be noted in Fig.~\ref{fig22}(b) is the dielectric anisotropy effect. It is seen that for all solvent densities, the transverse dielectric permittivity is larger than the longitudinal one, i.e. $\epa>\epe$. The dielectric anisotropy in the slit pore is a direct consequence of image dipole  interactions. Since the negative and positive elementary charges on the edges of each solvent molecule are subject to the same image forces, image dipole interactions favor the polarization of solvent molecules along the membrane surface. This results in a transverse permittivity that exceeds the longitudinal component.  Moreover, the inset of Fig.~\ref{fig22}(a) shows that the dielectric anisotropy monotonically increases with the bulk solvent density. This is again due to the amplification of the dielectric contrast between the membrane and the solvent with increasing solvent concentration. Then, in Fig.~\ref{fig22}(b), one notices that at large solvent concentrations where the dipolar alignment along the membrane surface becomes the strongest, the transverse permittivity exceeds the bulk permittivity, while the longitudinal component stays below the bulk value at all concentrations. Thus, for solvents at biological concentrations confined in nanoscale pores, the relevant correlation effect associated with the confinement is the anisotropic dielectric response of the liquid.  As noted above, the breaking of the bulk dielectric isotropy in nanoslits with the transverse permittivity exceeding the longitudinal and bulk permittivities has been already observed in MD simulations with explicit solvent~\cite{hansim}.  Our formalism allows the first unambigous interpretation of these simulation results in terms of the interfacial solvent correlations.

\begin{figure}
\includegraphics[width=.9\linewidth]{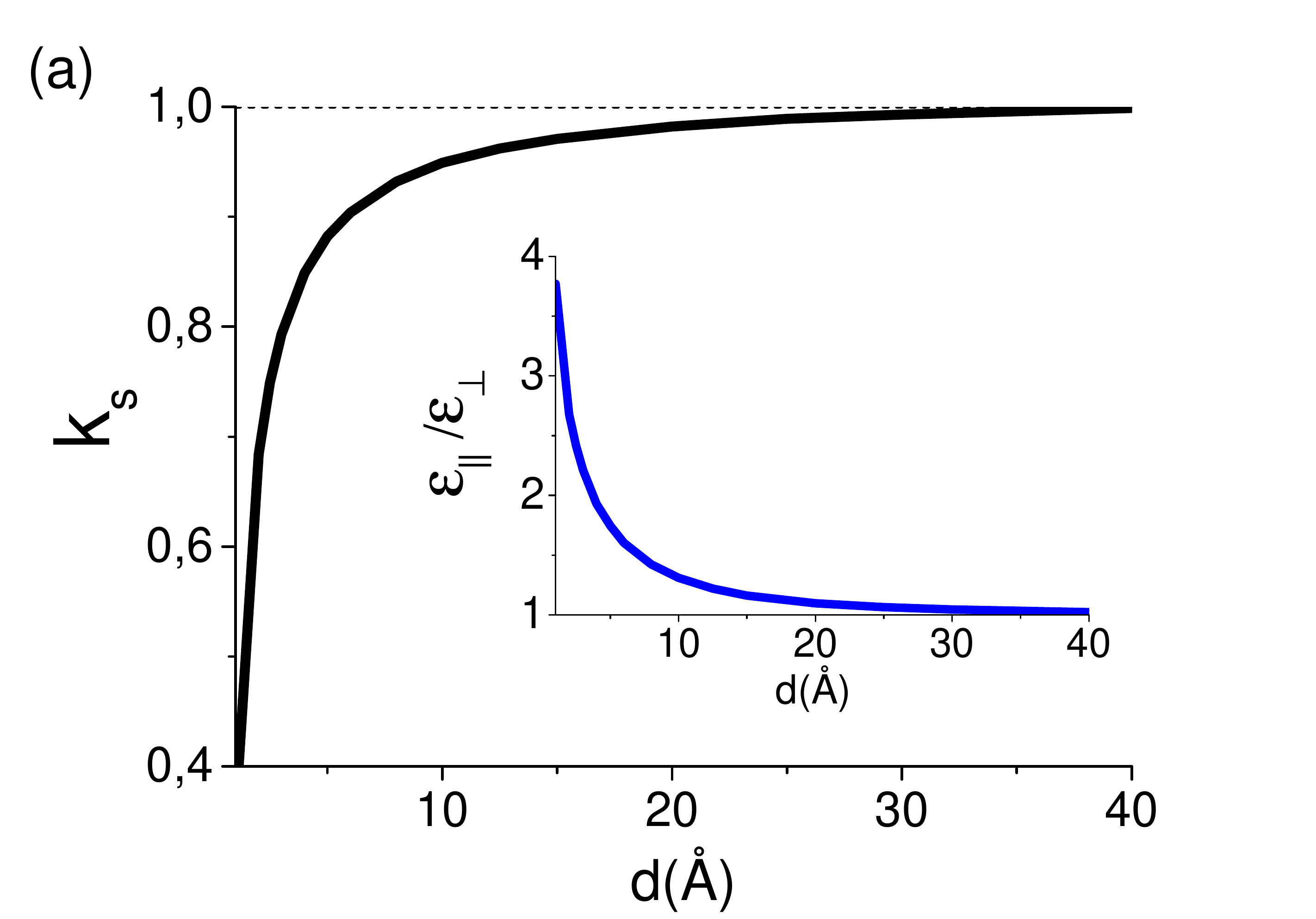}
\includegraphics[width=.9\linewidth]{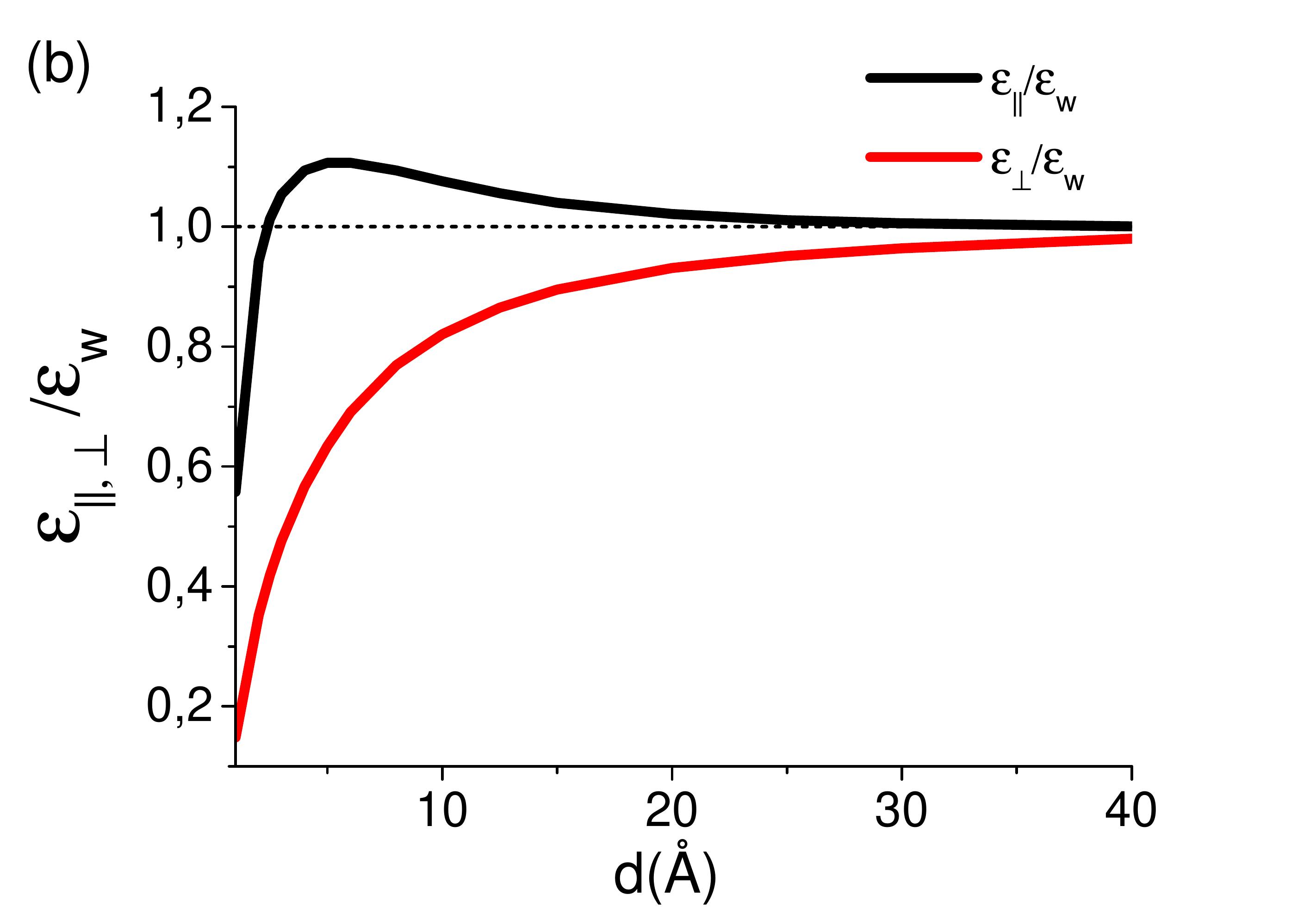}
\caption{(Color online)  (a) Solvent partition coefficient~(\ref{sopa}), (b) dielectric permittivity components renormalized by the reservoir permittivity ($\rho_{sb}=55$ M, and free of ions $\rho_{ib}=0.0$ M). The inset in (a) displays the ratio of the parallel and longitudinal dielectric permittivities.}
\label{fig23}
\end{figure}

Further, we consider the effect of the pore size on the behaviour of confined solvents at the physiological density $\rho_{sb}=55$ M. Figs.~\ref{fig23}(a) and (b) illustrate the solvent partition coefficient and the dielectric permittivities against the pore size varied between $d=1$ {\AA} and $40$ {\AA}.  As a consequence of the amplification of image dipole interactions with confinement, the solvent exclusion and the resulting dielectric reduction effects become relevant mostly at subnanometer pore sizes. Moreover, as the pore size is decreased, one sees that the dielectric reduction effect occurs in a more pronounced fashion for the longitudinal than the transverse component. The former decreases monotonically until it drops to $\epe\simeq 0.2\;\e_w$  for the pore thickness $d=1$ {\AA}. This is due to the combined effects of dipolar alignment and exclusion that are both amplified with decreasing pore size. However, because the dipolar alignment increases the transverse permittivity, we see that the latter first rises with decreasing pore size and then starts to drop below the bulk permittivity below a characteristic pore thickness $d\simeq 5$ {\AA} where the dipolar exclusion dominates the alignment effect. We also report in the inset of Fig.~\ref{fig23}(a) the ratio $\epa/\epe$ against the pore size. This plot shows that the dielectric anisotropy becomes very large for subnanometer pores, with the transverse permittivity being about four times larger than the longitudinal component at the pore size $d=1$ {\AA}. 

\subsubsection{Effect of solvent confinement on vdW interactions}
\label{vandw}

Van der Waals (vdW) interactions between low dielectric bodies play a major role in determining the stability of large colloids~\cite{isr}.  The standard formulation of vdW forces considers the solvent surrounding the colloids as a dielectric continuum liquid whose dielectric permittivity is unaffected by the presence of the macromolecules. However, we have shown that this assumption is incorrect for subnanometer intermolecular distances comparable to the solvent molecular size. Motivated by this point, we evaluate the impact of the dielectric reduction and anisotropy on the interaction between the neutral walls. The inner interaction force between the interfaces is given by the derivative of the total free energy~(\ref{vargr1}) with respect to the interplate distance, that is $\Pi_{in}(d)=-\delta\Omega_v/(S\delta d)$. Because we would like to make a qualitative comparison with the standard vdW forces \textcolor{black}{that takes into account only the quadratic fluctuations of the electrostatic field}, we will consider exclusively the vdW part of the free energy corresponding to the last term of Eq.~(\ref{om0ani}). Taking the limit of infinite cut-off $\Lambda\to\infty$, carrying out the integral in Fourier space and differentiating the vdW free energy with respect to the pore size $d$, the vdW level interaction force follows in the form
\be\label{vdw}
\Pi_{vdW}=-\frac{A}{8\pi d^3},
\ee
where we introduced an effective Hamaker constant
\be\label{ha}
A=\frac{\epe}{\epa}Li_3\left(\Delta_\gamma^2\right).
\ee
In Eq.~(\ref{ha}), $Li_3(x)$ is the polylogarithm function of third order~\cite{math}. The standard vdW interaction of the dielectric continuum formulation is also given by Eq.~(\ref{vdw}) but with a different Hamaker constant $A\to A_0=Li_3(\Delta_0^2)$ recovered in the limit $\e_{\pa,\perp}\to\e_w$, with the coefficient $\Delta_0$ given by Eq.~(\ref{diju})~\cite{isr}. 

\begin{figure}
\includegraphics[width=.9\linewidth]{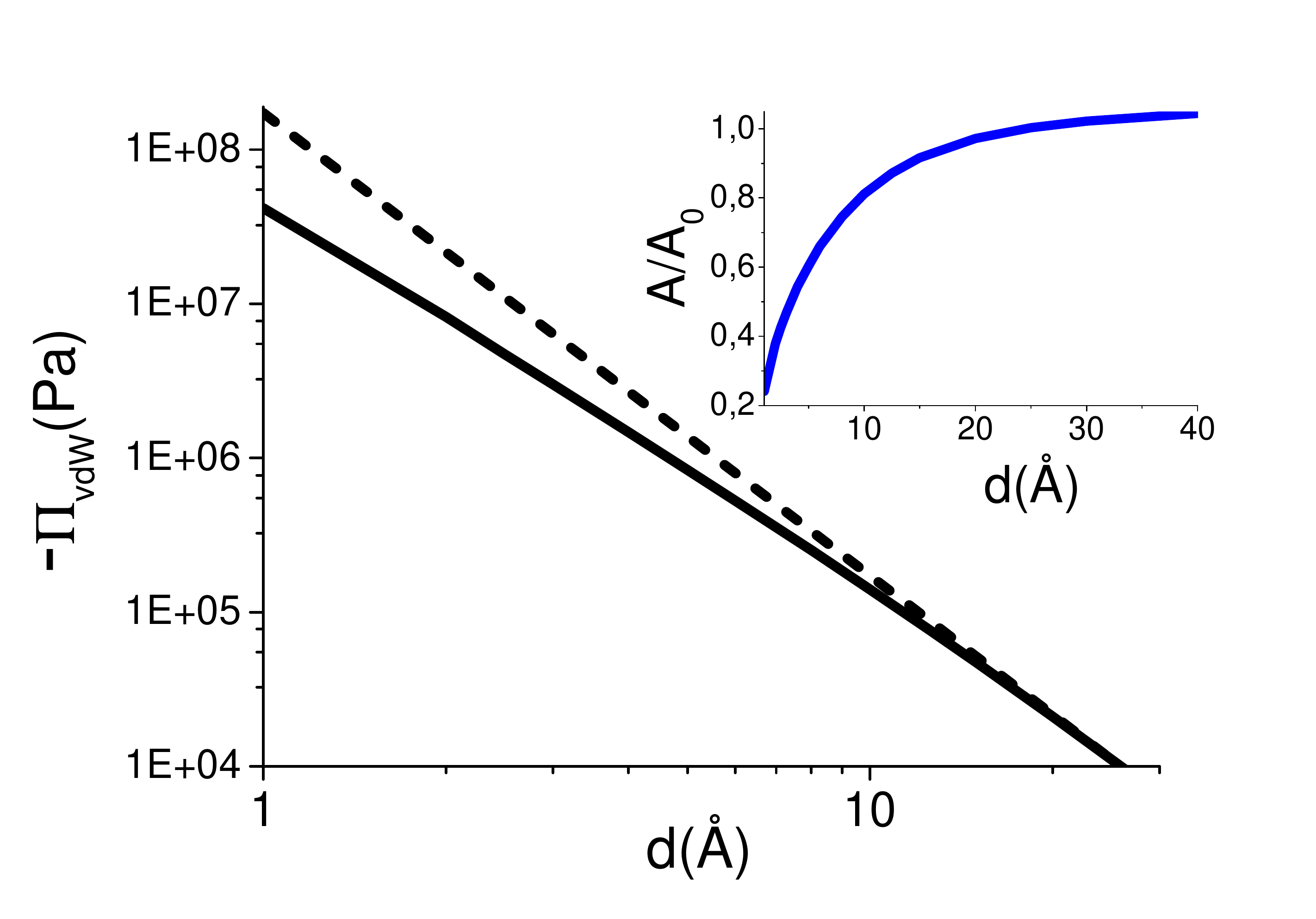}
\caption{(Color online)  The van der Waals part of the interplate pressure against the slit size for the solvent of bulk density $\rho_{sb}=55$ M and free of ions ($\rho_{ib}=0.0$ M). The inset shows the rescaled Hamaker constant.}
\label{fig24}
\end{figure}

Fig.~\ref{fig24} compares the interaction force in Eq.~(\ref{vdw}) (solid curve) with the usual vdW pressure (dashed curve). It is seen that for interplate separations in the range $d\lesssim1$ nm, the present theory predicts a significantly \textcolor{black}{less attractive} pressure than the vdW theory. To understand this point, we note that the dielectric anisotropy and dipolar exclusion effects result in the inequalities  $\epe/\epa<1$ and $\Delta_\gamma<\Delta_0$, respectively. According to Eq.~(\ref{ha}), this reduces the effective Hamaker constant below the standard value, that is $A<A_0$.  This is illustrated in the inset of Fig.~\ref{fig24} where the effective Hamaker constant is shown to be reduced for $d\lesssim1$ nm.  This indicates that for intercolloidal distances in the subnanometer range, the dielectric continuum formulation of macromolecular interactions neglecting solvent-membrane interactions overestimates the vdW attraction between low dielectric bodies. 

\subsection{Interfacial dielectric response and differential capacitances : local versus non-local correlations}
\label{cap}

In this last part we consider the effect of non-local correlations on the interfacial dielectric response of the polar solvent, and the differential capacitance of the electrolyte in contact with a charged plane located at $z=0$. The symmetric electrolyte is composed of two species of monovalent anions and cations. The single interface system follows from the slit geometry of Fig.~\ref{fig1} by taking the limit of infinite separation $d\to\infty$. In Section~\ref{dia}, it was shown that the dielectric reduction and anisotropy effects vanish in this limit. Thus, the results of this part will be obtained by solving Eq.~(\ref{varnlpb2}) with the ionic and dipolar self energies $\delta v_i(z)$ and $\delta v_d(z,a_z)$ approximated with the local DH-potential. The numerical solution scheme based on a relaxation algorithm and the explicit form of the self-energies are described in Appendix~\ref{relax}. In order to make comparisons with previous local results of the EDPB formalism~\cite{epl}, the bulk  solvent concentration will be set to the value $\rho_{sb}=50.8$ M, which is slightly lower than the density of water. The calculations were carried out with an infinite UV cut-off $\Lambda\to\infty$.
\begin{figure}
\includegraphics[width=1\linewidth]{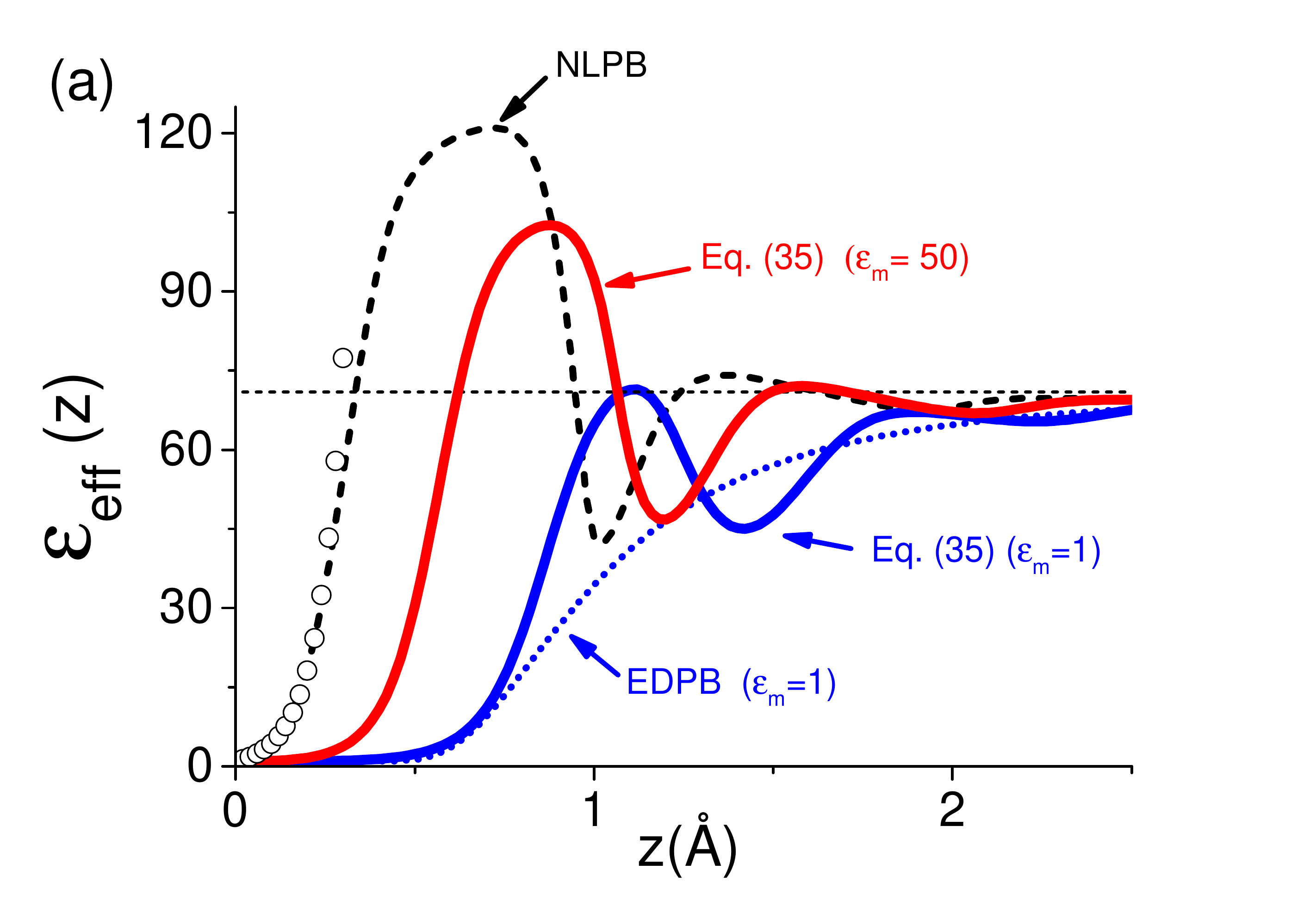}
\includegraphics[width=1\linewidth]{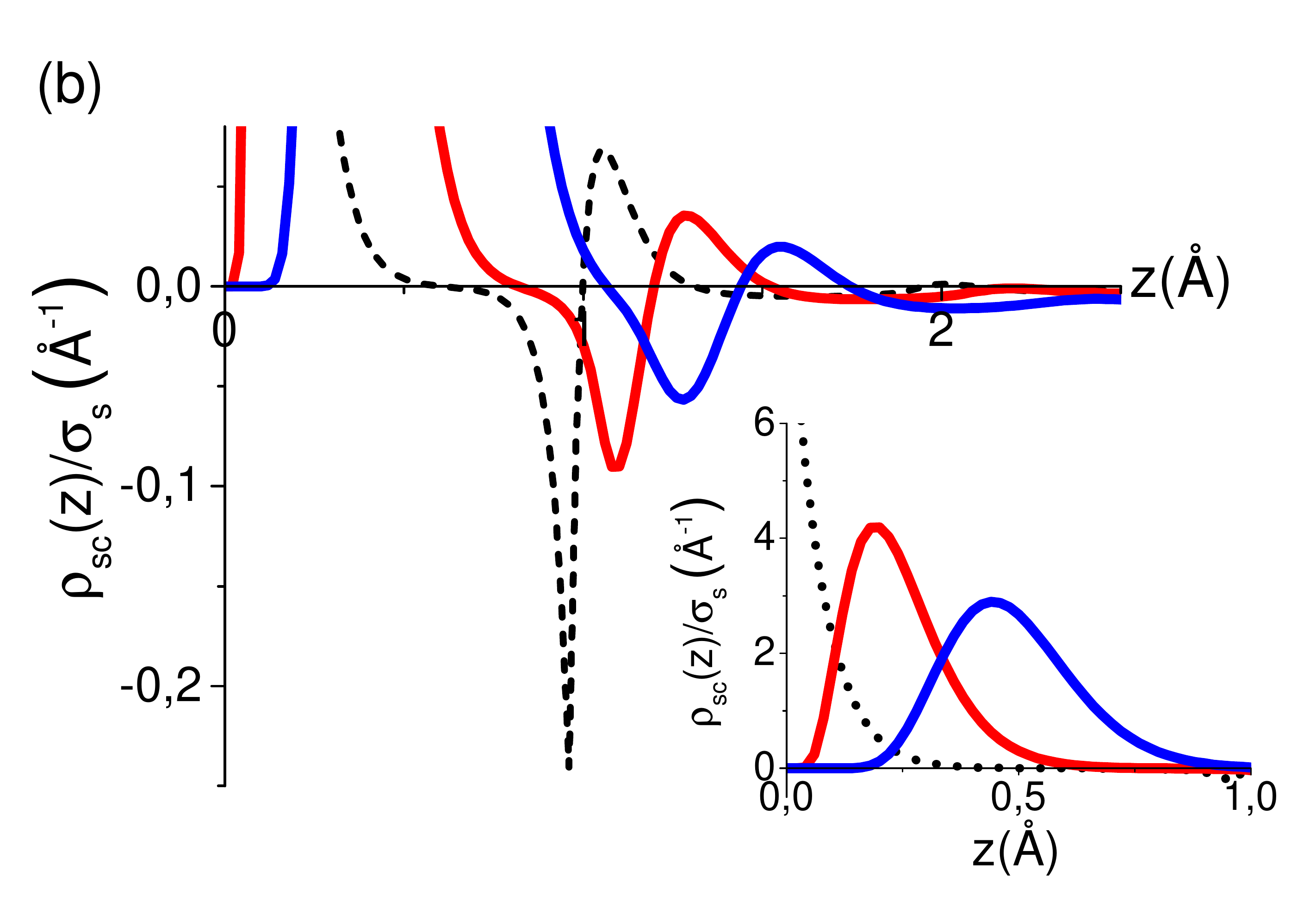}
\caption{(Color online)  (a) Effective dielectric permittivity profile and (b) rescaled polarization charge density obtained from Eq.~(\ref{varnlpb2}) for the solvent with bulk density $\rho_{sb}=50.8$ M and permittivity $\e_w=71$, in contact with a planar interface with surface charge $\sigma_s=10^{-6}$ e $\mbox{nm}^{-2}$. Salt concentration is $\rho_{ib}=10^{-5}$ M. The MF result (NLPB) reached in the limit $\e_m=\e_w$ is denoted by the dashed black curves, and the effect of correlations associated with the dielectric inhomogeneity between the solvent and the membrane is shown by the solid curves. The open circles in (a) correspond to the asymptotic limit of Eq.~(\ref{dilst}). In the same plot, we also show the permittivity profile of the point-dipole liquid model (EDPB-dotted blue curve) for which non-local effects are absent~\cite{epl}.}
\label{fig3}
\end{figure}

The electrostatic potential profiles $E(z)=\phi'_0(z)$ obtained from the solution of Eq.~(\ref{varnlpb2})  were used in order to obtain the effective permittivity of the liquid from the relation~\cite{NLPB1}
\be\label{e}
E(z)=\frac{e^2\sigma_s}{k_BT\e_{eff}(z)}.
\ee
We restrict ourselves to the linear response regime and eliminate ionic screening effects, we consider a very weak surface charge density $\sigma_s=10^{-6}$ e $\mbox{nm}^{-2}$ and a dilute salt with bulk concentration $\rho_{ib}=10^{-5}$ M.  The numerical result for the dielectric permittivity profile is shown in Fig.~\ref{fig3}(a) for various membrane permittivities. At the MF level, or equivalently for the dielectrically homogeneous system $\e_m=\e_w$ where one is left exclusively with non-local dielectric response effects, it is seen that the interface is characterized by a dielectric void followed by large oscillations of the effective permittivity function around the bulk permittivity.  In Ref.~\cite{NLPB1} where we treated the polar liquid model in MF approximation, it was shown that the oscillations of the permittivity result from the presence of successive hydration layers with alternating net charge at the interface. In order to extend this relation beyond the MF limit, we note that by integrating  Eq.~(\ref{varnlpb2}) from the interface to a given point $z$ in the liquid, and using relation~(\ref{e}), the inverse effective permittivity can be expressed in terms of the cumulative polarization charge between the surface and the point $z$, 
\be
\label{efpol}
\frac{1}{\e_{eff}(z)}=1-\frac{1}{\sigma_s}\int_0^z\mathrm{d}z'\rho_{sc}(z'),
\ee
with the solvent charge density given by $\rho_{sc}(z)=Q\left[\rho_{s+}(z)-\rho_{s-}(z)\right]$. Eq.~(\ref{efpol}) was derived in Ref.~\cite{NLPB1} for the MF NLPB model. This relation states that the trend of the effective permittivity should be reversed at the points where the polarization charge density changes its sign. This is illustrated in Fig.~\ref{fig3}(b). It is seen that for all membrane permittivities, the extrema of the effective permittivity correspond exactly to the boundary between two neighboring hydration shells of opposite charge.  Furthermore, in Fig.~\ref{fig3}(a), one notes that for membranes with low permittivity $\e_m<\e_w$, electrostatic correlations increase the size of the dielectric void close to the charged surface, shift the permittivity curves towards larger distances from the interface, and reduce the overall amplitude of the dielectric oscillations. With the aid of Eq.~(\ref{efpol}), these aspects can be again explained in terms of the solvent partition at the interface. Namely, in the main plot and the inset of Fig.~\ref{fig3}(b), we show that in the vicinity of low dielectric membranes where solvent molecules are subject to strong image forces, the polarization charges are shifted far away from the interface and their spatial variation is smoothed. 

Fig.~\ref{fig3}(a) also shows the dielectric permittivity profile of the point-dipole model developed in Ref.~\cite{epl}. The result illustrated by the dotted blue curve is obtained from the solution of the EDPB equation introduced in the same article.  We note that the EDPB formalism incorporates the interfacial solvent exclusion but not the non-local dielectric response. The comparison of the local (dashed blue curve) and non-local result (solid blue curve) shows that both models are characterised by a similar dielectric permittivity reduction at the interface, and the non-locality of the present solvent model manifests itself by the oscillatory behaviour of the permittivity curve around the local result. Thus, for low dielectric membranes associated with pronounced image forces, the resulting solvent depletion largely dominates the effect of non-locality and brings the main contribution to the interfacial dielectric reduction. 

\begin{figure}
\includegraphics[width=.9\linewidth]{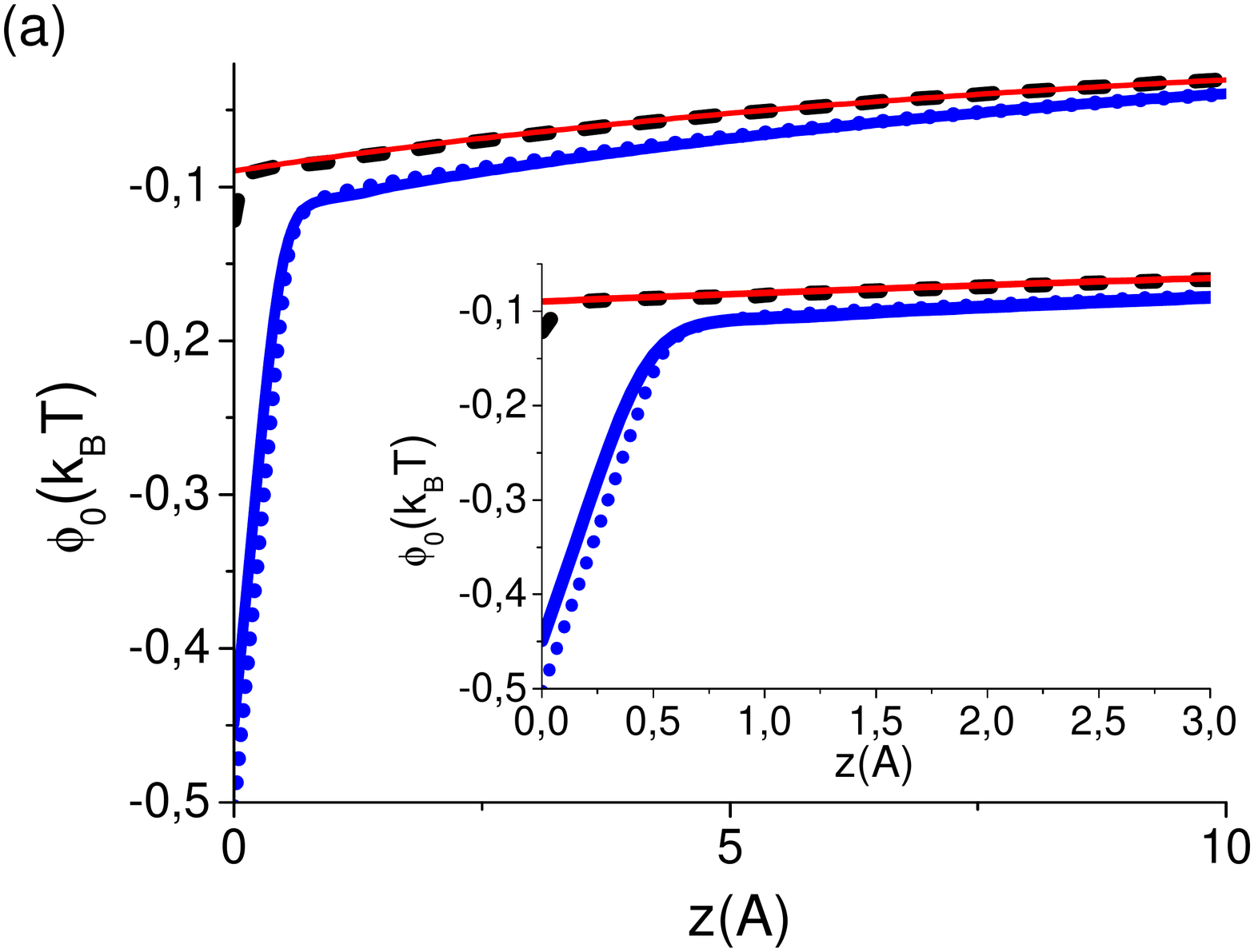}
\includegraphics[width=.9\linewidth]{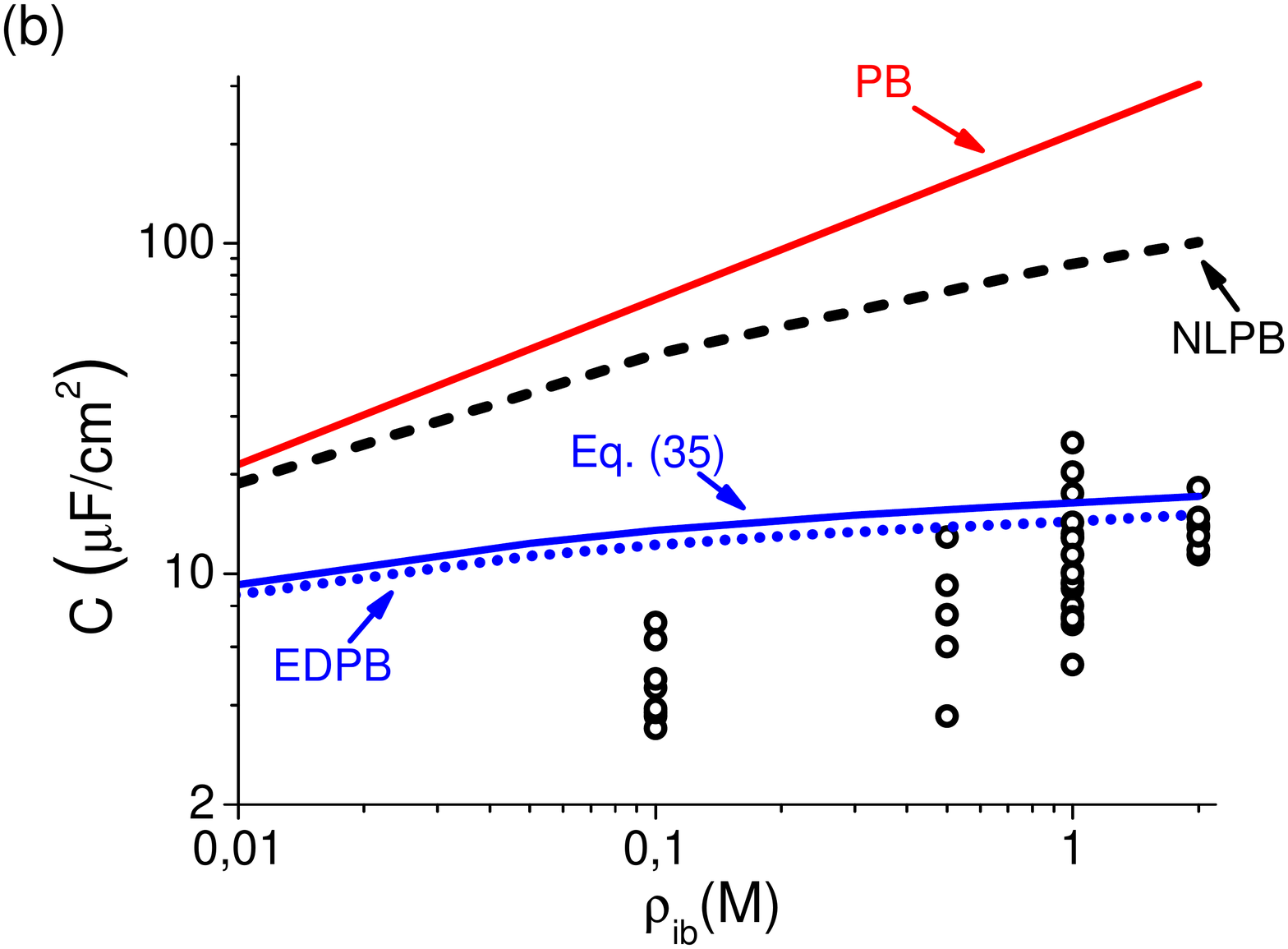}
\caption{(Color online)  (a) Electrostatic potential profile and (b) differential capacitance against the bulk salt concentration for the solvent with reservoir density $\rho_{sb}=50.8$ M in contact with a planar interface, with the membrane permittivity $\e_m=1$. In (a), the interfacial charge is $\sigma_s=0.01$ e $\mbox{nm}^{-2}$ and the salt density $\rho_{ib}=0.1$ M. The capacitances in (b) correspond to the point-of-zero-charge reached in the limit $\sigma_s\to0.0$. In both figures, the red curve is the PB result, and the other curves have the same annotation as in Fig.~\ref{fig3}. The open squares in (b) are the experimental capacitance data from Ref.~\cite{netzprl}.}
\label{fig4}
\end{figure}

We will now illustrate the impact of this point on the differential capacitance of low dielectric materials. 
The accurate knowledge of the dielectric response of water in the vicinity of low polarity substrates is crucial for the optimization of new generation energy devices such as supercapacitors, which are usually fabricated from carbon based materials with low static permittivities on the order $\e_m\simeq 2-5$, hence they are ideal realizations of the limit $\e_m \ll \e_w$. The standard theory that allows to predict the charge storage ability of these systems is the Gouy-Chapman (GC) model, which is based on the PB formalism~\cite{rev3}. Within the framework of this MF theory, the differential capacitance is defined as
\be
\label{dif}
C=\frac{q_ie^2}{k_BT}\left|\frac{\partial\sigma_s}{\partial\phi_0(0)}\right|
\ee
and is given by the simple formula $C=\e_w\kappa_{DH}$, where the DH parameter is defined in Eq.~(\ref{dh}).  It is well-known that the GC-capacitance overestimates by several factors the experimental capacitance data of materials at the point-of-zero-charge. This is illustrated in Fig.~\ref{fig4}(b) where the GC capacitance against the salt density is compared with experimental data taken from Ref.~\cite{netzprl}. In the same article where the authors used  in the PB equation the dielectric permittivities extracted from MD simulations, it was argued that the overestimation of the experimental capacitances by the GC theory is due to the absence of non-local effects in the latter. Then, within the local point dipole model of Ref.~\cite{epl}, we showed that the solvent exclusion in the vicinity of the capacitor electrode is sufficient to reproduce the low values of the experimental capacitance data (compare the dotted blue curve with the open circles in Fig.~\ref{fig4}(b)). 

To evaluate the importance of non-locality on the differential capacitance, we first compare in Fig.~\ref{fig4}(a) the potential profiles obtained from the MF PB and NLPB equations. It is seen that as a result of the thin interfacial dielectric reduction layer in Fig.~\ref{fig3}(a) (dashed black curve), the NLPB potential at the surface is slightly below the PB result. In Fig.~\ref{fig4}(b), one notices that this in turn lowers the PB capacitance, an effect that becomes more pronounced with increasing salt concentration but remains insufficient to explain the experimental trend. Then, in Fig.~\ref{fig4}(a), the electrostatic potential obtained from  the solution of Eq.~(\ref{varnlpb2}) is seen to be lower than the NLPB result by several factors, remaining very close to the result of the local EDPB formalism. This stems again from the dielectric screening deficiency associated with the large dielectric exclusion layer in Fig.~\ref{fig3}(a) (solid blue curve). Finally, in Fig.~\ref{fig4}(b), the same interfacial dielectric reduction amplified by image dipole interactions is seen to drop the differential capacitance of the MF NLPB formalism to the order of magnitude of the experimental data, improving over the GC capacitance by several factors. However, the comparison of the non-local and EDPB results indicates that non-locality brings a minor contribution to the capacitance. This point confirms that the interfacial solvent depletion induced by image interactions plays the leading role in the low amplitudes of the differential capacitances at the point-of-zero-charge.

\textcolor{black}{Finally, we evaluated the effect of non-local correlations on the average orientation of solvent molecules close to the charged interface. In order to compare the result of the present finite-size dipole model with the point dipole theory of Ref.~\cite{epl}, we have to derive the number density of solvent molecules with respect to their centers of mass at the point $\br+\ba/2$, where $\br$ denotes the position of the positive charge on the solvent molecule (see Fig.~\ref{fig1}). To this aim, we redefine the single-particle potential in the free energy of Eq.~(\ref{varf}) as $W_s(\br,\ba)=W_+(\br)+W_-(\br+\ba)+W_{cen}(\br+\ba/2)$, where the third potential acts on the center of the dipole. Shifting in Eq.~(\ref{varf}) the integration variable as $\br=\bf{u}-\ba/2$, evaluating the density with the relation $\rho_{d}(\br)=\delta \Omega_v/\delta W_{cen}(\br)$, and considering the planar symmetry of the system, one gets the dipolar number density in the form
\be\label{dencen}
\rho_d(z)=\int_{-a_i(z)}^{a_i(z)}\frac{\mathrm{d}a_z}{2a}f_d(z,a_z),
\ee
with the auxiliary function $a_i(z)=\mathrm{min}(a,2z)$ and the dipolar number density at fixed orientation
\bea
f_d(z,a_z)&=&\rho_{sb}\;e^{Q\left[\phi_0(z+a_z/2)-\phi_0(z-a_z/2)\right]}\nonumber\\
&&\hspace{.35cm}\times e^{-\frac{Q^2}{2}\delta v_d(z-a_z/2,a_z)}.
\eea
In terms of the dipolar number density in Eq.~(\ref{dencen}), the average orientation of dipoles can be evaluated by computing the local fluctuations of the dipole moment $p_z=Qa_z$ according to the relation
\be\label{avdip}
\mu_m(z)=\frac{\lan p_z^2\ran}{p_0^2\rho_d(z)},
\ee
where
\be
\lan p_z^2\ran=\int_{-a_i(z)}^{a_i(z)}\frac{\mathrm{d}a_z}{2a}f_d(z,a_z)a_z^2.
\ee
We note that freely rotating dipoles correspond to the value $\mu_m(z)=1/3$, and dipolar alignment along the membrane surface and perpendicular to the interface are respectively characterized by the regimes $\mu_m(z)<1/3$ and $\mu_m(z)>1/3$~\cite{Kanduc,bohdip,epl}.} 

\begin{figure}
\includegraphics[width=1.05\linewidth]{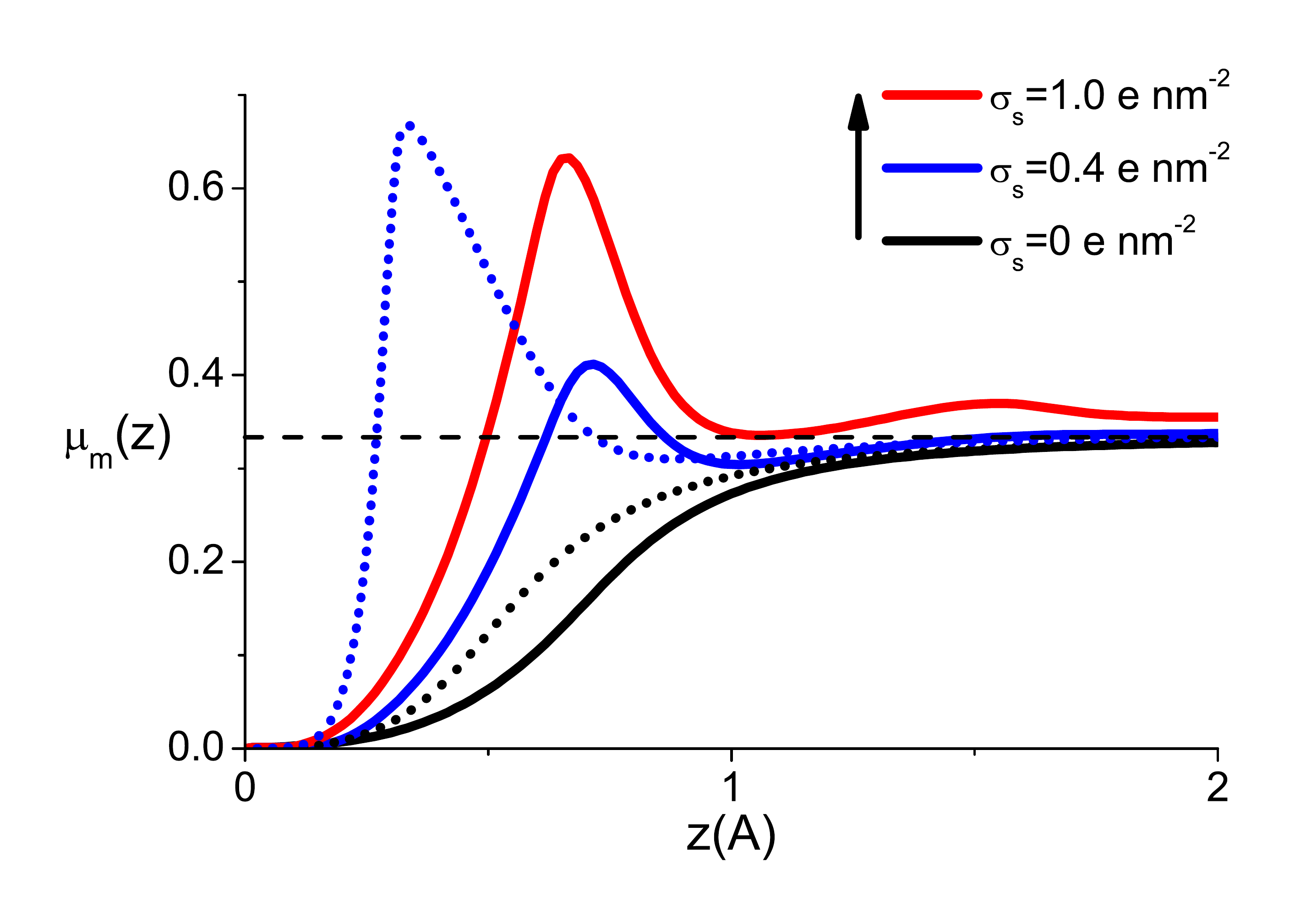}
\caption{\textcolor{black}{(Color online)  Average orientation of solvent molecules from Eq.~(\ref{avdip}) for the same parameters as in Fig.~\ref{fig4}(a) and various surface charges. The colours corresponding to each surface charge are given in the legend for finite size dipoles (solid curves) and point-dipoles (dotted curves from Ref.~\cite{epl}). The horizontal curve marks the freely rotating dipole regime with $\mu_m(z)=1/3$.}}
\label{fig5}
\end{figure}

\textcolor{black}{
The spatial variation of the order parameter $\mu_m(z)$ computed with the solution of Eq.~(\ref{varnlpb2}) is reported in Fig.~\ref{fig5} for various surface charges (solid curves). It is seen that close to the interface where image dipole interactions dominate dipole-surface charge interactions, solvent molecules exhibit an alignment parallel to the surface for all surface charge values. For neutral interfaces, moving from the surface towards the bulk region, this alignment is gradually replaced by the free rotation of dipoles. In the case of charged interfaces, at separation distances $z\gtrsim 0.5$ {\AA} where the surface charge-dipole interactions take over the image dipole forces, the alignment regime is replaced by the polarization of solvent molecules parallel to the external field, i.e. perpendicular to the surface. This second regime corresponds qualitatively to the MF behaviour of dipoles observed in previous dipolar PB theories~\cite{bohdip,Kanduc}. For $z\gtrsim 1$ {\AA} where surface-dipole interactions weaken, one tends to the freely rotating dipole case with $\mu_m(z)\simeq 1/3$.} 

\textcolor{black}{
In order to illustrate the importance of the finite dipole size on the dipolar orientation, we also reported in Fig.~\ref{fig5} the average rotation of point dipoles from Ref.~\cite{epl}. The comparison of the dashed and solid black curves shows that at neutral interfaces, as a result of the rotational penalty and stronger image-dipole interactions for finite size dipoles, the latter exhibit a stronger dipolar alignment than point dipoles. Then, at charged interfaces where surface charge-dipole interactions are weaker for finite size dipoles, point-dipoles show a stronger polarization along the external field. This indicates that in the point-dipole approximation, solvent-membrane interactions are underestimated at neutral surfaces but overestimated at charged interfaces.}

\section{Conclusions}

In this work, we investigated electrostatic fluctuations in inhomogeneously partitioned solvents with internal structure.  In Section~\ref{genvar}, we derived the non-local self-consistent equations~(\ref{varnlpb1}) and~(\ref{var2}) for the charged liquid in an arbitrary geometry. These equations which take into account the extended charge structure of the solvent molecules generalize the local variational equations of Netz and Orland from Ref.~\cite{netzvar}. In Section~\ref{dilsol}, the non-local equations were first solved at one-loop order for a dilute solvent confined in a nanoslit. We showed that the explicit consideration of the solvent at one-loop level already results in ionic image-charge interactions responsible for ion rejection from membrane nanopores~\cite{yar1}.  Furthermore, the difference between the solvent-explicit and local image-charge potentials was shown to result from the additional ionic Born energy between the pore and the ion reservoir. We also found that the latter becomes important exclusively for slits of subnanometric size. This observation allows to fix the confinement scale where the dielectric continuum electrostatics is expected to be valid.

In Section~\ref{dia}, we considered the solvent model at physiological concentrations within a restricted variational approch based on a dielectrically anisotropic kernel. It was found that the interaction of solvent molecules with their electrostatic images results in two effects. First of all, the dipolar alignment along the membrane surface leads to an anisotropic dielectric response of the fluid characterized by a transverse permittivity exceeding the longitudinal permittivity, i.e. $\epa>\epe$. Secondly, the solvent experiences a partial rejection from the nanopore, which in turn reduces the dielectric permittivity components below the reservoir value. We note that the same dielectric reduction effect has been observed in recent ion rejection experiments~\cite{lang}. Furthermore, the dielectric reduction and anisotropy become relevant for pores of subnanometric size, and both effects were shown to weaken the Hamaker constant of the standard vdW theory. This result indicates that the dielectric continuum formulation of macromolecular interactions overestimates the vdW attraction between dielectric bodies at small separation distances.

Section~\ref{cap} dealt with correlation effects on the interfacial dielectric response of the solvent at simple charged planes. We showed that the major effect of correlations is the decrease of the MF surface polarization charge densities, resulting in a reduced dielectric permittivity in the vicinity of the charged interface. The interfacial dielectric void induced by dipole-image interactions was shown to largely dominate the MF dielectric reduction associated with non-locality. We also compared with experimental capacitance data the prediction of the present non-local theory with and without correlations. We found that the MF theory including only solvent structure is not sufficient to explain the low values of the experimental data, and the consideration of correlations responsible for the strong interfacial solvent depletion is necessary to have a quantitative agreement with experiments. 

Being a first beyond-mean-field theory of inhomogeneous charged fluids with structured solvent, the present formalism naturally needs refinements. The most obvious limitation of the model is the absence of hard-core interactions between solvent molecules. The theory can be improved in this direction by incorporating steric effects as in Ref.~\cite{orland1} or considering repulsive Yukawa interactions between the particles~\cite{jstat}.  The absence of hydrogen-bonding in the present theory is an additional complication. Considering that hydrogen-bonding partially results from electrostatic interactions, one may have to include the second order cumulant corrections to the present self-consistent theory in order to cover this effect. Furthermore, the quantitative accuracy of our approach should of course be tested in simulations, which we plan to do first for bulk solvents in an externally applied electric field. Before concluding, we emphasize that recent ion transport studies revealed without ambiguity the importance of the inhomogeneous solvent partition on the elektrokinetic properties of charged liquids~\cite{netznano1,netznano2}. Thus, the formalism that we propose is susceptible to find important applications not only in artificial nanofiltration studies but also in nanofluidic techniques where the interpretation of experimental data is still based on the MF dielectric continuum electrostatics.
\\

{\bf Acknowledgement.} SB gratefully acknowledges support under the ANR blanc grant ``Fluctuations in Structured Coulomb Fluids''.

\smallskip
\appendix
\section{Solution of the non-local Laplace equation for dilute solvents}
\label{ap1}

In this Appendix, we explain the numerical solution of Eq.~(\ref{lap}) for dilute solvents. Defining the auxiliary function
\bea
F(z,z')&=&2Q^2\rho_{sb}\int_{a_1(z)}^{a_2(z)}\frac{\mathrm{d}a_z}{2a}\left\{\tv_0(z,z')\right.\\
&&\left.\hspace{3cm}-\tv_0(z+a_z,z')J_0(k|a_\pa|)\right\},\nonumber
\eea
and the reference kernel corresponding to the DH-kernel in vacuum
\be
\tv_r^{-1}(z,z')=-\frac{k_BT}{e^2}\left[\partial_z\e_0(z)\partial_z-\e_0(z)p^2_b\right]\delta(z-z'),
\ee
and making use of Eq.~(\ref{defgr}), the relation~(\ref{lap}) can be formally inverted as
\be\label{lapin}
\tv_0(z,z')=\tv_r(z,z')-\int_0^d\mathrm{d}z_1\tv_r(z,z_1)F(z_1,z').
\ee
The derivation of the reference DH-potential in vacuum is similar to the computation of the same potential in the dielectric continuum solvent, solution of Eq.~(\ref{dielcon}) (see e.g. Ref.~\cite{netzvdw} for details). It is given by the sum of a bulk and an interfacial part,
\be
\label{dhair}
\tv_r(z,z')=\tv_{rb}(z,z')+\delta\tv_r(z,z'),
\ee
where the homogeneous part reads
\be\label{dhairB}
\tv_{rb}(z,z')=\frac{2\pi\ell_B}{p_b}e^{-p_b|z-z'|},
\ee
and the contribution from the pore confinement is given by
\bea\label{dhairD}
\delta\tv_r(z,z')&=&\frac{2\pi\ell_B}{p_b}\frac{\Delta}{1-\Delta^2e^{-2p_bd}}\\
&&\hspace{0.5cm}\times\left\{e^{-p_b(z+z')}+e^{-p_b(2d-z-z')}\right.\nonumber\\
&&\hspace{1cm}\left.+2\Delta e^{-2p_bd}\cosh\left(p|z-z'|\right)\right\},\nonumber
\eea
with the auxiliary function $\Delta=(p_b-k)/(p_b+k)$. Eq.~(\ref{lapin}) allows to compute the non-local potential $\tv_0(z,z')$ by iteration around the reference potential~(\ref{dhair}).  Before concluding, it is useful to note that the DH potential $\tv_{DH}(z,z')$ in the dielectric continuum solvent can be recovered from Eq.~(\ref{dhair}) if one replaces the coefficients $\ell_B$,  $p_b$, and $\Delta$ in Eqs.~(\ref{dhairB})-(\ref{dhairD}) respectively by the Bjerrum length in the solvent 
\be\label{an1}
\ell_w=\frac{\ell_B}{\e_w},
\ee
the screening function
\be\label{an2}
\bar p=\sqrt{k^2+\kappa_{DH}^2},
\ee
and the dielectric jump function 
\be\label{an3}
\bar\Delta=\frac{\e_w\bar p-\e_mk}{\e_w\bar p+\e_mk}.
\ee

\section{Derivation of the electrostatic propagator with dielectric anisotropy}
\label{grani}

This Appendix introduces the dielectrically anisotropic Green's function that solves Eq.~(\ref{anila}). By injecting into this equation the Fourier expansion~(\ref{four1}), the former takes the one dimensional form
\be\label{anila2}
\left[-\partial_z\epe(z)\partial_z+\epa(z)k^2\right]\tv_0(z,z')=\frac{e^2}{k_BT}\delta(z-z'),
\ee
where the dielectric permittivity components $\e_{\pa,\perp}(z)$ are introduced in Eq.~(\ref{dielcom}). We note that in the present work, we need the solution of Eq.~(\ref{anila2}) exclusively for ions located in the slit, i.e. for $0\leq z'\leq d$. The general solution of Eq.~(\ref{anila2}) is then given by
\bea\label{gensol}
\tv_0(z,z')&=&C_1e^{kz}\theta(-z)+C_2e^{-kz}\theta(z-d)\\
&&+\left[C_3e^{\gamma kz}+C_4e^{-\gamma kz}\right]\theta(z'-z)\theta(z)\theta(d-z)\nonumber\\
&&+\left[C_5e^{\gamma kz}+C_6e^{-\gamma kz}\right]\theta(z-z')\theta(z)\theta(d-z),\nonumber
\eea
where we introduced the coefficient
\be\label{gma}
\gamma=\sqrt\frac{\epa}{\epe}.
\ee
Furthermore, the constants $C_i$ with $i=1,...,6$ in Eq.~(\ref{gensol}) are obtained from the continuity of the electrostatic potential $\tv_0(z,z')$ and the displacement field $\epe(z)\partial_z\tv_0(z,z')$ at the pore walls and at the location of the charge source $z=z'$. After some algebra, one gets the Fourier-transformed propagator in the form
\be
\label{dhw}
\tv_0(z,z')=\tv_{0b}(z,z')+\delta\tv_0(z,z'),
\ee
with the bulk part
\be
\label{dhwB}
\tv_{0b}(z,z')=\frac{2\pi\ell_B}{k\sqrt{\epe\epa}}e^{-\gamma k|z-z'|}
\ee
and the dielectric part
\bea\label{dhwD}
\delta\tv_0(z,z')&=&\frac{2\pi\ell_B}{k\sqrt{\epe\epa}}\frac{\Delta_\gamma}{1-\Delta_\gamma^2e^{-2\gamma kd}}\\
&&\times\left\{e^{-\gamma k(z+z')}+e^{-\gamma k(2d-z-z')}\right.\nonumber\\
&&\hspace{5mm}\left.+2\Delta_\gamma e^{-2\gamma kd}\cosh\left(\gamma k|z-z'|\right)\right\},\nonumber
\eea
where we defined the dielectric discontinuity function
\be\label{dlta}
\Delta_\gamma=\frac{\sqrt{\epe\epa}-\e_m}{\sqrt{\epe\epa}+\e_m}.
\ee

\section{Computation of the anisotropic vdW free energy}
\label{vdwder}

We present in this Appendix the derivation of the vdW part of the variational grand potential given by
\be\label{om0}
\Omega_0=-\ln\int \mathcal{D}\phi\;e^{-\frac{k_BT}{2e^2}\int\mathrm{d}\br\left\{\epa(z)\left[\left(\partial_x\phi\right)^2+\left(\partial_y\phi\right)^2\right]+\epe(z)\left(\partial_z\phi\right)^2\right\}}.
\ee
In order to evaluate the functional integral, we will employ the charging procedure introduced in Ref.~\cite{netzvdw} for the computation of the isotropic vdW energy. Our approach consists in reexpressing the relation~(\ref{om0}) in terms of two auxiliary integrals over the charging parameters $\xi$ and $\eta$,
\begin{widetext}
\bea\label{om02}
\Omega_0&=&-\int_0^1\mathrm{d}\xi\frac{\mathrm{d}}{\mathrm{d}\xi}\ln\int \mathcal{D}\phi\;e^{-\frac{k_BT}{2e^2}\int\mathrm{d}\br\left\{\e_\xi(z)\left[\left(\partial_x\phi\right)^2+\left(\partial_y\phi\right)^2\right]+\epe(z)\left(\partial_z\phi\right)^2\right\}}\nonumber\\
&&-\int_0^1\mathrm{d}\eta\frac{\mathrm{d}}{\mathrm{d}\eta}\ln\int \mathcal{D}\phi \;e^{-\frac{k_BT}{2e^2}\int\mathrm{d}\br\;\e_\eta(z)\left(\nabla_\br\phi\right)^2}-\ln\int \mathcal{D}\phi \;e^{-\frac{k_BT}{2e^2}\int\mathrm{d}\br\;\e_m\left(\nabla\phi\right)^2},
\eea
\end{widetext}
where we introduced the auxiliary permittivity functions 
\bea\label{e1}
\e_\xi(z)&=&\epe(z)+\xi\left[\epa(z)-\epe(z)\right] \\
\label{e2}
\e_\eta(z)&=&\e_m+\eta\left[\epe(z)-\e_m\right].
\eea
We now note that the third integral on the r.h.s. of Eq.~(\ref{om02}) is simply the free energy of a bulk medium with dielectric permittivity $\e_m$, and this contribution is independent of the variational parameters $\epe$, $\epa$, and the pore size $d$.  Thus, in the following part of the derivation, this constant will be dropped. Evaluating  the derivatives acting on the functional integral in Eq.~(\ref{om02}), the free energy takes the form
\begin{widetext}
\bea\label{om03}
\Omega_0=\frac{k_BT}{2e^2}\int\mathrm{d}\br\left\{\int_0^1\mathrm{d}\xi\left[\epa(z)-\epe(z)\right]\lan\left(\nabla_{\br_\pa}\phi\right)^2\ran_{\epa(z)\to\e_\xi(z)}
+\int_0^1\mathrm{d}\eta\left[\epe(z)-\e_m\right]\lan\left(\nabla_\br\phi\right)^2\ran_{\e_{\pa,\perp}(z)\to\e_\eta(z)}\right\}.
\eea
\end{widetext}
In Eq.~(\ref{om03}), the subscripts of the brackets mean that the averages should be evaluated with the electrostatic Green's function~(\ref{dhw}) by replacing the dielectric permittivity profiles of the latter with the auxiliary permittivity functions~(\ref{e1}) and~(\ref{e2}),
\bea\label{v01}
v_0^\xi(\br,\br')&=&v_0\left[\br,\br';\epa(z)\to\e_\xi(z)\right]\\
\label{v02}
v_0^\eta(\br,\br')&=&v_0\left[\br,\br';\e_{\pa,\perp}(z)\to\e_\eta(z)\right].
\eea
Evaluating the field-theoretic averages, the free energy~(\ref{om03}) can be expressed in terms of the propagators~(\ref{v01}) and~(\ref{v02}) as
\begin{widetext}
\bea
\label{om04}
\Omega_0&=&\frac{k_BT}{2e^2}\left.\int\mathrm{d}\br\left\{\int_0^1\mathrm{d}\xi\left[\epa(z)-\epe(z)\right]\nabla_{\br_\pa}\cdot\nabla_{\br'_\pa}v_0^\xi(\br,\br')+
\int_0^1\mathrm{d}\eta\left[\epe(z)-\e_m\right]\nabla_\br\cdot\nabla_{\br'}v_0^\eta(\br,\br')\right\}\right|_{\br'\to\br}\\
\label{om05}
&=&S\frac{k_BT}{4\pi e^2}\left.\int_0^d\mathrm{d}z\int_0^\Lambda\mathrm{d}kk\left\{\left(\e_\pa-\e_\perp\right)\int_0^1\mathrm{d}\xi\;k^2\tv_0^\xi(z,z')
+\left(\e_\perp-\e_m\right)\int_0^1\mathrm{d}\eta\left[k^2+\partial_z\partial_{z'}\right]\tv_0^\eta(z,z')\right\}\right|_{z'\to z},
\eea
\end{widetext}
where  $S$ stands for the lateral surface of the membrane, and the second equality follows from the first one after substituting the Fourier expansion of the electrostatic propagator Eq.~(\ref{four1}). Carrying out the integrals over the pore size and the auxiliary parameters in Eq.~(\ref{om05}), after some lengthy algebra, one gets the anisotropic vdW part of grand potential given by Eq.~(\ref{om0ani}) of the main text.

We finally note that in Eq.~(\ref{om05}), the successive derivatives acting on the bulk part of the Green's function (Eq.~(\ref{dhwB}) transformed according to Eq.~(\ref{v02})) yields a delta function evaluated at zero,
\be\label{derz}
\left.\partial_z\partial_{z'}\tv_{0b}^\eta(z,z')\right|_{z'\to z}=\frac{4\pi\ell_B}{\e_\eta}\delta(0),
\ee
which is indeed finite since the UV modes are regularized with a cut-off,
\be\label{del}
\delta(0)=\int_0^{\Lambda_z}\frac{dk_z}{2\pi}=\frac{\Lambda_z}{2\pi}.
\ee
The longitudinal cut-off $\Lambda_z$ can be related to the transverse one $\Lambda$ by noting that in the isotropic limit $\epe=\epa=\e_w$, the spherical symmetry should be recovered, that is
\bea\label{c1}
&&\left.\partial_z\partial_{z'} v_{0b}(\br-\br')\right|_{\br'\to\br}=\frac{\ell_B}{\e_w}\frac{\Lambda^2\Lambda_z}{2\pi}\\
&&=\left.\partial_x\partial_{x'} v_{0b}(\br-\br')\right|_{\br'\to\br}=\left.\partial_y\partial_{y'} v_{0b}(\br-\br')\right|_{\br'\to\br}\nonumber\\
\label{c2}
&&=\frac{\ell_B\Lambda^3}{6\e_w}.
\eea
From the equalities~(\ref{c1}) and~(\ref{c2}), one gets the relation $\Lambda_z=\pi\Lambda/3$, which gives with Eq.~(\ref{del}) $\delta(0)=\Lambda/6$. Using the latter in Eq.~(\ref{derz}), one finally obtains
\be\label{derz2}
\left.\partial_z\partial_{z'}\tv_{0b}^\eta(z,z')\right|_{z'\to z}=\frac{2\pi\ell_B\Lambda}{3\e_\eta}.
\ee

\section{Computation of the dipolar contribution to the grand potential}
\label{dipcon}

In this Appendix, we explain the derivation of the dipolar part~(\ref{dipom}) of the variational grand potential in Eq.~(\ref{vargr1}). After evaluating the field-theoretic average in Eq.~(\ref{var1}), one finds for the contribution from solvent molecules 
\be\label{od}
\Omega_d=-S\Lambda_s\int_0^d\mathrm{d}z\int_{a_1(z)}^{a_2(z)}\frac{\mathrm{d}a_z}{2a}e^{E_s-\frac{Q^2}{2}v_d(z,a_z)}.
\ee
In Eq.~(\ref{od}), the  dipolar self-energy is composed of a bulk and a surface contribution,
\bea\label{ditot}
v_d(z,a_z)=v_{db}(a_z)+v_{ds}(z,a_z),
\eea
where the bulk part corresponding to the Born energy of solvent molecules is
\be\label{dipb}
v_{db}(a_z)=\frac{2\ell_B\e_0}{\sqrt{\epa\epe}}\int_0^\Lambda\mathrm{d}k\left[1-e^{-\gamma k|a_z|}J_0(ka_\pa)\right],
\ee
and the inhomogeneous part accounting for the confinement reads
\bea\label{dips}
 v_{ds}(z,a_z)&=&\int_0^\Lambda\frac{\mathrm{d}kk}{2\pi}\left\{ \delta\tv_0(z,z)+ \delta\tv_0(z+a_z,z+a_z)\right.\nonumber\\
&&\hspace{1.5cm}\left.-2  \delta\tv_0(z,z+a_z)J_0(ka_\pa)\right\},
\eea
with the potential $\delta\tv_0(z,z')$ given by Eq.~(\ref{dhwD}). 

In order to obtain the relation between the solvent fugacity and the reservoir density, we have to compute the grand potential~(\ref{vargr1}) for a bulk solvent. First of all, we note that in the bulk limit $d\to\infty$ and $\Delta_\gamma=0$, the surface contribution~(\ref{dips}) naturally vanishes. Then, because we chose a local form~(\ref{anila}) for the reference electrostatic kernel, the bulk self-energy~(\ref{dipb}) has to be expanded in the point-dipole limit in order to recover the cut-off free bulk solution~(\ref{dibu}) for the effective permittivities. Taking the point-dipole limit of Eq.~(\ref{dipb}) that consists of its Taylor expansion up to the order $O(a^2)$, one gets
\be\label{dipb2}
v_{db}(a_z)=\frac{\ell_B\Lambda^3\e_0}{6\sqrt{\epa\epe}}\left(a_\pa^2+\gamma a_z^2\right).
\ee
In the same bulk limit $V=Sd\to\infty$, the grand potential per volume $f_b=\Omega_v/V$ reads 
\bea\label{ob}
f_b&=&\frac{\Lambda^3}{12\pi}\gamma+\frac{\Lambda^3}{16\pi}\ln\frac{\epe}{\e_m}+\frac{\Lambda^3}{48\pi}\left[\frac{2(\e_0-\epa)}{\sqrt{\epa\epe}}+\frac{\e_0}{\epe}\right]\nonumber\\
&&-\Lambda_s\frac{\sqrt\pi}{2}\frac{\mathrm{erf}(u)}{u}e^{-\alpha},
\eea
where we introduced the coefficients $\alpha=Q^2\Lambda^3\ell_Ba^2\e_0/(12\sqrt{\epa\epe})$ and $u=\sqrt{\alpha(\gamma-1)}$. From the numerical minimization of the bulk grand potential~(\ref{ob}) with respect to $\epa$ and $\epe$, we found that the permittivity components are given by $\epa=\epe=\e_w$, in agreement with our early result in Ref.~\cite{NLPB2}. Substituting this solution into Eq.~(\ref{ob}), and evaluating the solvent density with the thermodynamic relation $\rho_{sb}=-\Lambda_s\partial f_b/\partial\Lambda_s$, we obtain the relation between the solvent fugacity and density
\be\label{fugden}
\Lambda_s=\rho_{sb}\exp\left(\frac{Q^2\e_0\ell_Ba^2\Lambda^3}{12\e_w}\right).
\ee
Injecting into Eq.~(\ref{od}) the potential~(\ref{ditot}) with Eqs~(\ref{dips}) and~(\ref{dipb2}), and the fugacity~(\ref{fugden}), one finally obtains the dipolar term of the grand potential given by Eq.~(\ref{dipom}) of the main text.

\section{Numerical algorithm for the solution of the correlation corrected NLPB equation~(\ref{varnlpb2})}
\label{relax}

In this Appendix, we present a relaxation algorithm for the solution of the correlation corrected NLPB equation~(\ref{varnlpb2}). The electrolyte is composed of two species of monovalent cations and anions, that is $q_i=1$. To simplify the notation, we also set the valency of the charges on the solvent molecules to $Q=1$. In the single interface limit $d\to\infty$, Eq.~(\ref{varnlpb2}) takes the form
\bea\label{1in}
&&\partial_z^2\phi_0(z)-\kappa_{DH}^2\;e^{-\delta v_i(z)/2}\sinh\left[\phi_0(z)\right]\\
&&-\kappa_s^2\int_{a_1(z)}^a\frac{\mathrm{d}a_z}{2a}e^{-\delta v_d(z,a_z)/2}\sinh\left[\phi_0(z)-\phi_0(z+a_z)\right]\nonumber\\
&&=4\pi\ell_B\sigma_s\delta(z),\nonumber
\eea
with the dipolar screening function $\kappa_s^2=8\pi\ell_B\rho_{sb}$.  The ionic and dipolar self-energies in Eq.~(\ref{1in}) will be approximated with the DH potential $v_{DH}(\br,\br')$ whose derivation is explained at the end of Appendix~\ref{ap1}. In the single interface limit, these self-energies defined by Eqs.~(\ref{is2i}) and~(\ref{is2d})  read 
\bea
&&\delta v_i(z)=\ell_w\int_0^\infty\frac{\mathrm{d}kk}{\bar p}\bar\Delta e^{-2\bar p z}\\
&&\delta v_d(z,a_z)=\ell_w\int_0^\infty\frac{\mathrm{d}kk}{\bar p}\bar\Delta e^{-2\bar p z}\\
&&\hspace{3.4cm}\times\left[1+e^{-2\bar p a_z}-2e^{-\bar p a_z}J_0(ka_\pa)\right],\nonumber
\eea
with the coefficients $\ell_w$, $\bar p$, and $\bar\Delta$ introduced by Eqs.~(\ref{an1})-(\ref{an3}). The integro-differential equation~(\ref{1in}) should be solved with the boundary conditions 
\bea\label{infi}
&&\phi_0(z\to\infty)=0\\
\label{gauss}
&&\phi'_0(0)=4\pi\ell_B\sigma_s,
\eea
where Eq.~(\ref{gauss}) that corresponds to the Gauss' law in vacuum follows from the integration of Eq.~(\ref{1in}) in the vicinity of the boundary at $z=0$. 

We solve Eq.~(\ref{1in}) on a discrete lattice located between $z=0$ and $z=z_{max}=100\;\ell_B$. The lattice is composed of $2N+1$ mesh points separated by the distance $\epsilon$. We now define the potential on the lattice as $\psi_n\equiv\phi_0(z_n)$, where the index $n$ denoting the position on the lattice runs from $1$ to $2N+1$. We also define the discrete form of the distance from the interface as $z_n=(n-1)\epsilon$. By using the finite difference expression for the Laplacian, $\epsilon^2\phi_0''(z)=-2\psi_n+\psi_{n+1}+\psi_{n-1}$,  Eq.~(\ref{1in}) can be rearranged in the discrete form
\bea
\label{nlpbdis}
\psi_n&=&\frac{1}{2}\left\{\psi_{n+1}+\psi_{n-1}-r\;e^{-\delta v_i(z_n)/2}\sinh\left(\psi_n\right)\right.\\
&&\left.-s\sum_{j=j_1(n)}^{j_2(n)}e^{-\delta v_d(z_n,a_z\to z_{j-n+1})/2}\sinh\left(\psi_n-\psi_j\right)\right\},\nonumber
\eea
with the coefficients $r=\epsilon^2\kappa_{DH}^2$, $s=\epsilon^3\kappa_s^2/(2a)$, and the auxiliary functions $j_1(n)=\mathrm{max}(1,n-n_a+1)$, $j_2(n)=n+n_a-1$, where the index $n_a$ is defined as $z_{n_a}=a$. Eq.~(\ref{nlpbdis}) should be coupled with the Gauss' law~(\ref{gauss}) that reads in discrete space $\psi_0=\psi_1-4\pi q_i\ell_B\sigma_s\epsilon$. 

The relaxation method consists in solving Eq.~(\ref{nlpbdis}) by iteration around a reference potential $\phi_r(z)$ whose choice is tricky. Indeed, this reference potential has to satisfy the same Gauss' law~(\ref{gauss}) as Eq.~(\ref{1in}). As the usual PB equation obeys a different Gauss' law, namely $\phi'_0(0)=4\pi\ell_w\sigma_s$, the iteration of Eq.~(\ref{nlpbdis}) around the PB potential profile does not converge. Thus, we will derive the reference potential by considering the asymptotic small distance limit of the linear MF NLPB equation. The latter is obtained from Eq.~(\ref{1in}) by neglecting the ionic and dipolar self energies and linearizing the equation in the potential $\phi_0(z)$. In the close neighborhood of the interface at $z=0$, the linear MF NLPB equation takes for $z\ll a$ and $\kappa_{DH}^2\ll\kappa_s^2$ the simple asymptotic form $\phi''_0(z)-\kappa_s^2\phi(z)/2=4\pi\ell_B\sigma_s\delta(z)$. Integrating this equation once, one finds for the electric field 
\be\label{asym}
\phi'_0(z)=4\pi\ell_B\sigma_se^{-\kappa_sz/\sqrt2}.
\ee
\\
We now note that the electrostatic field in Eq.~(\ref{asym}) is also the solution of the non-uniform Poisson equation $\partial_z\e(z)\partial_z\phi(z)=4\pi\ell_B\sigma_s\delta(z)$ characterized by the local permittivity function
\be\label{dilst}
\e(z)=e^{\kappa_sz/\sqrt{2}}.
\ee
In Fig.~\ref{fig3}(a), the asymptotic law~(\ref{dilst}) is shown to accurately reproduce the rise of the MF dielectric permittivity in the vicinity of the interface at $0\leq z\lesssim d_1$ (see open circles), with the characteristic distance $d_1$ defined by $e^{\kappa_sd_1/\sqrt2}=\e_w$. We thus choose the reference potential as the solution of the non-linear PB equation for $z\geq d_1$, and as the integral of the the asymptotic form~(\ref{asym}) for $z\leq d_1$. Imposing the countinuity of the potential at $z=d_1$, the reference potential follows in the piecewise form
\bea\label{refp}
\phi_r(z)&=&-4\;\mathrm{arctanh}\left(te^{-\kappa_{DH}z}\right)\theta(z-d_1)\\
&&\left\{\frac{4\pi\ell_w\sigma_s\sqrt2}{\kappa_s}\left[1-e^{-\kappa_s(d_1-z)/\sqrt2}\right]\right.\nonumber\\
&&\left.-4\;\mathrm{arctanh}\left(te^{-\kappa_{DH}d_1}\right)\right\}\theta(d_1-z),\nonumber
\eea
with the auxiliary parameter $t=\sqrt{1+(\kappa_{DH}\mu)^2}-(\kappa_{DH}\mu)$ and the Gouy-Chapman length $\mu=1/(2\pi\ell_w\sigma_s)$. At the first iterative level, one has to inject the reference potential~(\ref{refp}) into the r.h.s. of Eq.~(\ref{nlpbdis}), and the updated potential profile $\left\{\psi_n\right\}_n$ is to be used at the next iterative step as the input function. This cycle is continued until the potential profile is stabilized.

\end{document}